\documentclass[]{jfm}  %arxiv no row line

\usepackage{graphicx}
\usepackage{newtxtext}
\usepackage{newtxmath}
\usepackage{natbib}
\usepackage{hyperref}
\hypersetup{
    colorlinks = true,
    urlcolor   = blue,
    citecolor  = black,
}

\newcommand{\RomanNumeralCaps}[1]
\linenumbers
\usepackage{overpic}

\shorttitle{Analysis of high-energy drop impact onto deep liquid pool}
\shortauthor{H. Wang, S. Liu, A-C. Bayeul-Lainé, D. Murphy, J. Katz, O. Coutier-Delgosha}

\title{Analysis of high-energy drop impact onto deep \\liquid pool}

\author{Hui Wang\aff{1}\corresp{\email{hui.wang@ensam.eu}},
  Shuo Liu\aff{1},
  Annie-Claude Bayeul-Lainé\aff{1},
  David Murphy\aff{2},
  Joseph Katz\aff{3},
 \and Olivier Coutier-Delgosha\aff{1,4}\corresp{\email{ocoutier@vt.edu}}}
\affiliation{\aff{1}Univ. Lille, CNRS, ONERA, Arts et Metiers Institute of Technology, Centrale Lille, UMR 9014 - LMFL - Laboratoire de Mécanique des Fluides de Lille - Kampé de Fériet, F-59000 Lille, France
\aff{2}Department of Mechanical Engineering, University of South Florida, Tampa, FL 33620, USA
\aff{3}Department of Mechanical Engineering, Johns Hopkins University, 3400 N. Charles Street,
Baltimore, MD 21218, USA
\aff{4}Kevin T. Crofton Department of Aerospace and Ocean Engineering, Virginia Tech, Blacksburg, VA 24060, USA}

\begin{document}
\maketitle
%----------------------------------------------------------------%
%----------------------------------------------------------------%
\begin{abstract}
The present work is devoted to the analysis of drop impact onto a deep liquid pool. It is focused on the effects of high-energy splash regimes, caused by the impact of large raindrops at high velocities. Such cases are characterized by short time scales and complex mechanisms, and they have thus received little attention until now. The BASILISK open-source solver is used to perform three-dimensional Direct Numerical Simulations (DNS). The capabilities of the octree adaptive mesh refinement techniques enable to capture the small-scale features of the flow, while the Volume of Fluid (VOF) approach combined with a balanced force surface tension calculation is applied to advect the volume fraction of the liquids and reconstruct the interfaces. The numerical results compare well with experimental visualizations: both the evolution of crown and cavity, the emanation of ligaments, the formation of bubble canopy, and the growth of the downward spiral jet that pierces through the cavity bottom, are correctly reproduced. Reliable quantitative agreements are also obtained regarding the time evolution of rim positions, cavity dimensions and droplet distributions through an observation window. Furthermore, simulation gives access to various aspects of the internal flows (velocity, vorticity, pressure, energy budget), which allows to better explain the corresponding physical phenomena. Details of the early dynamics of bubble ring entrapment and splashing performance, the formation/collapse of bubble canopy, and the spreading of drop liquid, are discussed. The statistics of droplet size show the bimodal distribution in time, corroborating distinct primary mechanisms of droplet production at different stages.
\end{abstract}

%------------------1 Introduction---------------------------------%

\section{Introduction}
\label{sec:intro}

The impact of raindrops on a deep liquid pool has been extensively studied since the initial works of \citet{worthington1883impact, worthington1908study}. Various behaviours are obtained, depending primarily on the miscible or immiscible character of the two liquids \citep{lhuissier2013drop, castillo2016impact}, but also on their difference of density \citep{thomson1886v, manzello2002experimental, villermaux2022equilibrated}, the depth of the receiving volume \citep{macklin1969subsurface, wang2000splashing, fedorchenko2004some}, and the angle of the impact \citep{zhbankova1990collision, okawa2008effect, gielen2017oblique, liu2018experimental}, to mention only a few parameters.

In the specific case of identical liquids of density $\rho$, with a $90^\circ$ impact of the drop falling in the air on a deep volume of target liquid, very different phenomena are still observed, when the liquid viscosity and surface tension with air, the diameter $d$ of the drop and the speed of the impact $V$ are varied. \citet{schotland1960experimental} has first reported the primary effect of the Weber number $We=\rho V^{2}d/\sigma$ on the physics of the impact, i.e. the ratio of the drop kinetic energy to the energy required to deform the target liquid surface. 

A few years later, \citet{engel1966crater, engel1967initial} has provided a detailed description of various behaviours obtained from waterdrop impact on water pool when $We$ is varied in the range $3000\sim20000$. Based on visualizations of the impact and using white particles in the target liquid and red ink in the impacting drop, she has shown that the drop impact creates a cavity in the flat free surface, and subsequently the rise of the target liquid in a cylindrical shape, from the edges of the cavity. A bubble-thin cylindrical sheet of liquid is erected at the upper edge of this crown, which eventually necks in and closes in a bubble dome in the most energetic test cases. After that, the cavity shallows and a jet forms at the cavity floor, which flows through the centre of the crown in case it is open or merges with a downward jet coming from the top of the bubble dome in case the crown has closed. The author suggests that the cavity may vibrate at its maximal expansion if all the drop kinetic energy is not yet transformed into potential energy, and she also assumes that: (\romannumeral1) the maximum possible bubble height is equal to the cavity diameter, (\romannumeral2) the pressure evolution below the cavity floor should explain the formation of the upward jet, (\romannumeral3) the liquid of the initial drop is carried by this jet and will eventually form a secondary bubble at the centre of the cavity.

In the following decades, this analysis has been progressively completed by additional experiments with increasing visualization capabilities \citep{rodriguez1985some,pumphrey1989underwater, rein1996transitional}, leading to characteristic laws of cavity size and growth time \citep{leng2001splash} and an improved understanding of the phenomena mentioned hereabove \citep{rein1993phenomena}. Indeed, these studies were usually focused on a specific mechanism obtained in a small range of $We$ and Froude number $Fr=V^{2}/gd$, like for example the crater formation and collapse \citep{prosperetti1989underwater}, the break-up of the upward jet for $We$ between 5 and 200 \citep{manzello2002experimental, castillo2015droplet}, the bubble entrapment under the floor of the cavity and the subsequent rising thin jet \citep{thoroddsen2003air, deng2007role, hendrix2016universal}, the breakup of the crown rim \citep{deegan2008rayleigh}, and the air layer trapped below the impacting drop when the respective surfaces of liquid are deformed due to the local pressure changes at the early stage of the process \citep{tran2013air}.

In the most recent works, attention has been especially focused on these first stages of the impact, where the entrapped air layer under the drop is ruptured, resulting in the generation of secondary bubbles \citep{pumphrey1990entrainment, thoroddsen2003air, thoroddsen2012micro}. Immediately after the impact, the contact line between the drop and the receiving liquid, also called the “neck region”, moves radially at high speed. It results in a liquid sheet, composed of water coming from the target liquid, which is ejected horizontally outwards at the base of the contact surface \citep{weiss1999single, thoroddsen2002ejecta, josserand2003droplet}. This ejecta eventually breaks up into ligaments and droplets that result in the splashing of very fine sprays \citep{thoroddsen2011droplet, zhang2012splashing}. For a sufficiently high Reynolds number, this phenomenon comes earlier than the sheet-like jet that is directly produced by the drop impact and will later generate the crown formation, as observed by \citet{zhang2012evolution} using fast X-ray phase contrast imaging. The combination of these two jets was analysed by \citet{agbaglah2015drop} who used a combination of X-ray imaging and axisymmetric simulations to discuss the interaction between the two jets.

One important output is that this interaction may be strongly related to instabilities observed in the neck region. Indeed, the base of the ejecta may become unstable and generate some alternate vortex shedding in the liquid below the neck region \citep{thoraval2012karman, castrejon2012experimental}. A strong interplay between this von K\'{a}rm\'{a}n vortex street and the inception/break-up of the ejecta sheet is suggested by several recent studies \citep{agbaglah2015drop, li2018early}, but the mechanisms are still not elucidated. Non-axisymmetric behaviours have been observed in the vortex shedding and the ejecta formation processes from the bottom visualizations in cases of impact on thin water film \citep{,thoraval2013drop, li2018early}, which shows that these phenomena break the axisymmetry at the fine scale, and may be related to three-dimensional mechanisms. Local streamwise vortices, generated at the sharp corners at the basis of the ejecta at an early stage of the drop impact, are mentioned by these authors to possibly explain these non-axisymmetric effects. Using ultra-fast imaging, an early azimuthal instability was also detected by \citet{li2018early}, which consists of small azimuthal waves that grow at the edge of the outer line of the contact, before the inception of the shedding. This clearly poses a new challenge to numerical studies, which mostly assume axisymmetry \citep{thoraval2012karman, agbaglah2015drop}. The authors point out the need of three-dimensional simulations with sufficient resolutions of the finest scale to explore the vortex structure in the neck region and the interaction between the two corners of the base of the ejecta sheet, which may be a primary mechanism of this instability. 

As for the later stages after impact, phenomena occurring specifically at high Weber numbers, like the formation of ligaments at the edges of the crown, the ejection of droplets, the closure of the upper rim, and the subsequent downward jet, have been also investigated by \citet{engel1966crater}, \citet{deegan2007complexities}, \citet{bisighini2010crater}, \citet{ castillo2015droplet}, and most recently by \citet{murphy2015splash}, who provided detailed information about the length of the ligament according to time, and the population of aerosolized droplets. 

\begin{figure}
  \centerline{\includegraphics[width=\linewidth]{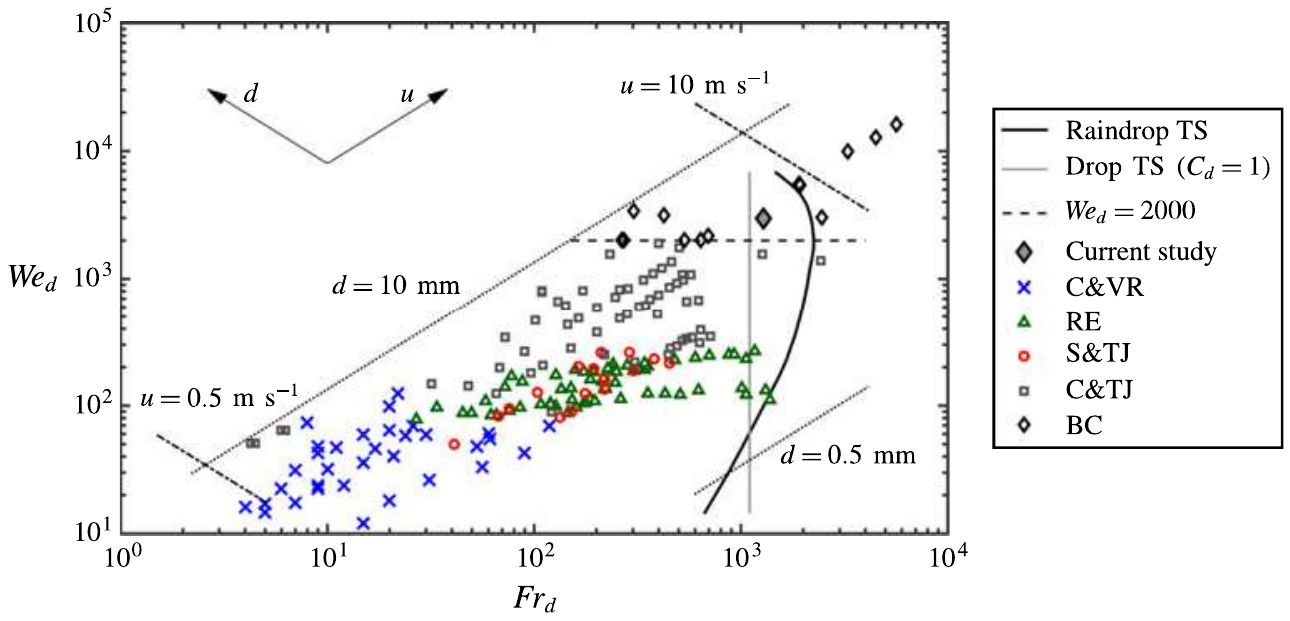}}% Images in 100% size
  \caption{Various configurations of splashing reported in the literature according to the $We$ and $Fr$ numbers reproduced from \citet{murphy2015splash} with the authorization of the authors.}
\label{fig:regime}
\end{figure}

An exhaustive review of these previous works devoted to the impact of a drop on a deep volume of the same low-viscosity liquid has been conducted by \citet{murphy2015splash}, as shown in figure \ref{fig:regime}. Five different regimes have been identified by authors, depending on the values of the Froude and Weber numbers. The two directions of variations of the impact speed $V$ and the drop diameter $d$ are plotted in figure \ref{fig:regime}, as well as the specific case of raindrops of different sizes falling at terminal speed (solid black line labelled ``Raindrop TS").

As can be seen in figure \ref{fig:regime}, if the drop size and the impact speed increase (moving typically from the bottom left to the top right on the chart), the following successive regimes are obtained: (1) slow drop that coalesces with the liquid volume and generates a vortex ring that moves downwards as the drop sinks, (2) formation of a cavity together with a surface wave, (3) the crater formed by increasing the drop diameter and the surface wave forms a crown rim from which the secondary droplets and ligaments are ejected, (4) for the highest speeds and the largest raindrop diameters, a large cavity is obtained, around which extends vertically a very thin liquid film, whose elevation and radius keep increasing and it eventually collapses in a very short time. At the upper part of the film, droplets and ligaments are ejected upwards and outwards. 

This last type of behaviour (black diamonds in figure \ref{fig:regime}) is the most energetic and only received very limited attention during the past twenty years \citep{pan2008collision, bisighini2010crater, sochan2018shape, lherm2021rayleigh}. This high-energy regime is characterized by shorter time scales and increased complexity of the phenomena involved in the splashing, which makes both experimental and numerical approaches more challenging, and may explain why it has received less attention until now. However, it is the case likely to produce the greatest number of aerosol droplets. The present study is focused on this highly energetic case of large raindrops falling near terminal speed, which is representative of raindrops at the surface of the ocean.

This configuration has been studied by \citet{murphy2015splash} in detail, primarily as a reference case for a study focused on the influence of oil slicks and oil dispersants on the impact. High-speed videos have enabled to characterise the time evolution of the external shape of the cavity, the crown and its upper rim, while microscopic holography has provided some statistics of droplet population ejected at different stages of the process. These previous experiments are used to validate the numerical strategies conducted in the present study in section \ref{sec:validation}, which is focused on the same conditions of drop diameter and speed at the impact. Further attention is given here to the early-time bubble ring entrapment and splashing near the impact neck, the internal mechanisms of crown formation as well as its violent closure, the liquid jets formed successively upward and downward and the subsequent entrainment of a large air bubble, and the ejection of droplets at the edge of the upper rim, in various directions that depend on the stage of the process. Indeed, it was estimated in the experiments that such a large raindrop produces at least 2000 micro-droplets. However, the details of the production of these tiny droplets and statistics on their size, speed, and ejection direction remain to be found, which motivated the use of numerical simulations in this paper.

Multiphase flow calculations of drop impacts usually consist in laminar simulations, as the Reynolds number $Re=\rho Vd/\mu$ based on the drop motion in the air at the impact velocity is below $10^4$, so there is no transition to turbulence, and all motions resulting from the impact are characterized by short characteristic times that do not involve any turbulence effect. The challenge of such computations is thus mostly related to the prediction of the multiphase structure. Early numerical works \citep{harlow1967splash, ferreira1985computer} based on Marker and Cell and SOLA-VOF techniques of interface tracking could provide some first predictions of the main features of drop impact but could not resolve the small-scale mechanisms. After that, \citet{oguz1990bubble} have included the surface tension effects and \citet{morton2000investigation} have solved the full Navier-Stokes equations, opening the way to modern simulations that are based either on interface tracking (level set, front tracking) or interface reconstruction, using a transport equation for the volume fractions of one component (VOF method).

The recent simulations of liquid drops impinging onto the liquid surface are mostly based on the latter category of models. Both axisymmetric simulations \citep{morton2000investigation, berberovic2009drop, ervik2014experimental, ray2015regimes, agbaglah2015drop, deka2017regime, fudge2021dipping} and three-dimensional simulations \citep{rieber1999numerical, brambilla2013automatic, cheng2015numerical, shin2017solver} can be found in literature, using mostly DNS approaches. For three-dimensional calculations, whose primary objective is to capture the non-axisymmetric mechanisms involved in the splashing, efforts are focused on dynamic refinement techniques \citep{nikolopoulos2007three, brambilla2015assessment}, in order to resolve the multiple interfaces resulting from the splashing, while ensuring a reasonable grid size. In the latter study, the authors show that a minimum cell size of about 50 to 100 $\upmu$m is sufficient to capture most of the features of the splash. However, that conclusion was drawn for Weber numbers around 250, so it may not directly apply to the high-energy configuration investigated in the present study.

The paper is organized as follows: the numerical methods and problem statement are described hereafter in section \ref{sec:methods}, the detailed validation of the numerical strategies is conducted by comparing the numerical results with the experimental data in section \ref{sec:validation}, the analysis of several mechanisms involved in the splashing is presented in section \ref{sec:overall}, and the statistics of airborne droplets is finally analyzed in section \ref{sec:droplet}.
%------------------2 Numerical method-----------------------------%

\section{Numerical methods and flow configurations}\label{sec:methods}

\subsection{Main features of the solver}\label{subsec:congif_approch}
The drop impact of the gas-liquid system is considered as incompressible flow, and it solves a system of the mass balance equation, the momentum balance equation and the advection of one-fluid formulation, called hereafter the colour function.
\begin{equation}
  \nabla\bcdot\boldsymbol{U}=0    
  \label{eq:mass}
\end{equation}
\begin{equation}
  \rho\left(\frac{\p \boldsymbol{U}}{\p t}+(\boldsymbol{U}\bcdot\nabla)\boldsymbol{U}\right)=-\nabla P+\nabla\bcdot(\mu\mathsfbi{D})+\rho\boldsymbol{a}+\sigma\kappa\delta_s\boldsymbol{n}
  \label{eq:momentum}
\end{equation}
\begin{equation}
  \frac{\p C}{\p t}+\nabla\bcdot(C\boldsymbol{U})=C\nabla\bcdot\boldsymbol{U}
  \label{eq:fraction}
\end{equation}
Where $\boldsymbol{U}$ is the velocity vector, $\rho$ is the density, $P$ is the pressure, $\mu$ is the viscosity, $\mathsfbi{D}$ is the deformation tensor whose components are $D_{ij}={\p u_{i}}/{\p x_{j}}+{\p u_{j}}/{\p x_{i}}$ with $u_i$ and $x_i$, $i=1$ to 3 the components of $\boldsymbol{U}$ and position $\boldsymbol{X}$, respectively. $\boldsymbol{a}$ is the body force along the impact direction. The last term in equation \ref{eq:momentum} is the surface tension force, with $\kappa$ the curvature of the interface, $\sigma$ the surface tension coefficient which is taken as constant in the present study, and the $\boldsymbol{n}$ the unit vector normal to the interface. This term is zero everywhere but at the gas/liquid interface, as controlled by the Dirac function $\delta_s$. The colour function $C$ is transported by equation \ref{eq:fraction}.  For incompressible flow, the right-hand term in equation \ref{eq:fraction} is zero due to equation \ref{eq:mass}. 

The BASILISK framework is a flow solver developed by \citet{popinet2015quadtree} and available for use and development under a free software GPL license \citep{basilisk}. It solves the time-dependent compressible/incompressible variable-density Euler, Stokes, or Navier-Stokes equations with second-order space and time accuracy. The Momentum-Conserving Volume-of-Fluid (MCVOF) approach implemented by \citet{fuster2018all} is employed here to simulate the problem of gas-liquid two-phase flow. The colour function \ref{eq:fraction} is solved based on a volume fraction advection scheme proposed by \citet{weymouth2010conservative}, which exhibits complete mass conservation and makes it ideal for highly energetic free-surface flows. The interface then can be represented using Piecewise Linear Interface Capturing (PLIC) VOF method \citep{rudman1998volume}. For equation \ref{eq:momentum}, the Crank–Nicholson discretization of the viscous terms is second-order accurate in time and unconditionally stable, while the convective terms are computed using the Bell-Colella-Glaz (BCG) second-order unsplit upwind scheme \citep{bell1989second}, which is stable for CFL numbers smaller than one. The calculation of the surface tension is one of the most challenging steps of the process, since no continuous definition of the interface is available in the VOF approach. Here the balanced-force surface-tension calculation is used \citep{francois2006balanced}, which is based on the Continuum-Surface-Force (CSF) approach originally proposed by \citet{brackbill1992continuum}. In addition, a second-order accurate calculation of the curvature is performed, using the Height-Function technique developed by \citet{popinet2009accurate}. More detailed descriptions of the numerical schemes can be found in  \citet{fuster2018all}, \citet{pairetti2020mesh} and \citet{zhang2020modeling}.

Cubic finite volumes organized hierarchically in octree are used for space discretization. The octree structure has been developed initially for image processing \citep{samet1990design} and later applied to CFD \citep{khokhlov1998fully} and multiphase flows \citep{popinet2003gerris}. The basic organization is the following, all details can be found in \citet{popinet2015quadtree}: when a cell is refined, it is divided into 8 cubic cells, whose edges are half the ones of the parent cell. The base of the tree is the root cell and the last cells with no child are the leaf cells. The cell level is the number of times it is refined, compared with the root cell (level 0). To avoid too much complexity in the gradient and flux calculations, the levels of direct and diagonally neighbouring cells are constrained and cannot differ by more than one and all cells directly neighbouring a mixed cell must be at the same level. 

BASILISK employs the Adaptive Mesh Refinement (AMR) \citep{van2018towards} to adaptively refine/coarsen the grids based on the wavelet-estimated numerical error of the local dynamics, which makes it especially appropriate for the present application where multiple interfaces and numerous droplets and bubbles are expected. The resolution will be adapted every time step according to the estimated discretisation error of the spatially-discretized fields (volume fraction, velocity, curvature, etc). The mesh will be refined as long as the wavelet-estimated error exceeds the given threshold, which eventually leads to a multi-level spatial resolution from the minimum refinement level $L_{min}$ to the maximum refinement level $L_{max}$ over the entire domain.

The BASILISK code has been already applied to the study of drop impact onto liquid film \citep{josserand2003droplet, reijers2019oblique, wu2021comparison, fudge2021dipping, sanjay2022drop} and the great superiority of parallel capacity and computational efficiency of the solver were discussed in \citep{wu2021comparison}. Furthermore, the capabilities of the BASILISK solver have been extensively validated on various problems of multi-phase flows here \citet{basilisk}.

\subsection{Initial Flow configurations}\label{subsec:congif_initial}
\begin{figure}
  \centering
  \begin{overpic}[width=\linewidth]{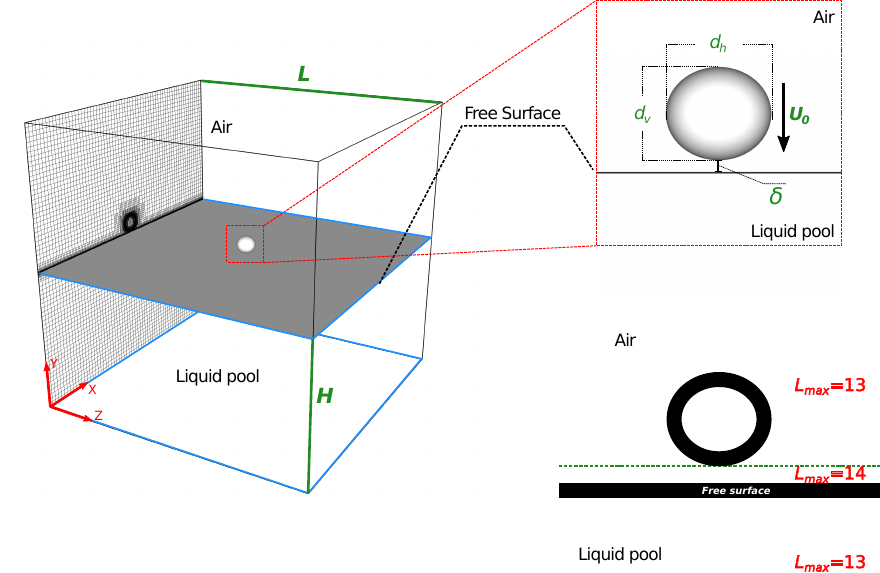} % Images in 100% size
  \put(2,55){\small(\textit{a})}
  \put(60,63){\small(\textit{b})}
  \put(60,30){\small(\textit{c})}
  \end{overpic}
  \caption{Initial numerical configuration of the three-dimensional simulation. (\textit{a}) Overall view of the computational domain and the initial mesh structure at a plane across the centre of the impacting drop (z = 0). (\textit{b}) Closeup view of the initial flows around the impacting drop. (\textit{c}) Mesh refinement strategy at the initial stage ($S1$). A higher maximum level of refinement ($L_{max}=14$) is imposed near the neck region to capture the early-time splashing.}
  \label{fig:initial}
\end{figure}

The configuration of drop impact investigated in the present study mimics the one studied by \citet{murphy2015splash}, using high-speed video. In the experiments, the drop falls in a $15.2\times15.2$ ${\rm cm^2}$ tank filled with seawater to a depth of 8 cm. The measured horizontal and vertical diameters of the drop just before the impact are $d_h=4.3$ mm and $d_v=3.8$ mm, respectively, which results in an effective drop diameter $d=(d_v d_h^2)^{1/3}=4.1$ mm. The ratio between the width of the tank $L$ and the drop diameter $d$ is therefore defined as $L=38d$. The speed before impact is $U_0=7.2$ m/s, which is 81\% of the drop terminal speed. 

In our numerical simulation, the computational domain is reduced down to a cube with a side length $L=16d$, which represents about 1/14 of the water tank volume in the experiments, as shown in figure \ref{fig:initial}(\textit{a}). It has been found to be the best compromise to avoid any effect of the boundary conditions on the splashing, while decreasing as much as possible the dimensions of the domain. The free surface is located at the mid-distance between the bottom and the top, thus the depth of the pool is $H=8d$, to give enough space to the aerosolized droplets without any interaction with the boundaries. The free outflow boundary condition is imposed on the top of the domain, while the default slip boundary condition (symmetry) is applied for the four side walls and the bottom. A zoomed-in view of the initial flow set-ups in the vicinity of the drop is depicted in figure \ref{fig:initial}(\textit{b}). The initial gap between drop and pool is $\delta=0.1d$, allowing the observation of air sheet entrainment near the contact line. The water in the drop and the tank are the same, which is assumed to have almost no effect on the splashing, as the properties of the freshwater drop and the target seawater in the experiments are very close. The density and viscosity ratios between seawater and air are  $\rho_w/\rho_a=1018$ and $\mu_w/\mu_a=180$, which leads to a system of drop-pool impact with dimensionless numbers $Re=28800$, $We=2893$ and $Fr=1322$.

As illustrated in figure \ref{fig:initial}(\textit{a}), the initial mesh configuration around the drop is generated based on the AMR algorithm using the estimated-discretization error of volume fraction ($f_{Err}=1e-6$) and velocity ($u_{Err}=1e-4$) fields, which promises a rounded geometric description of the initial drop and lowers the RAM requirement of initialization. The mesh is coarsened gradually down to the given minimum level of refinement ($L_{min}$) away from the drop interface. The region around the free surface of the pool is refined at $L_{max}=11$ to avoid any divergence issue. Once the simulation starts, the mesh will be redistributed adaptively based on AMR using the volume fraction field with tolerance $f_{Err}=1e-4$ and the velocity field with tolerance $u_{Err}=1e-1$ as adaption criteria. Additionally, we remove the droplets that approach the boundaries of the computational domain, since these tiny droplets have few effects on the evolution of the main impact but are expensive to track. In real experiments, they are considered to evaporate at one point and this is not our focus in this paper. Using the initial contact centre as the reference, any microdroplet whose centroid lies outside of the region of a semi-sphere with diameter $D_{remove}=15d$ will be removed from the simulation. The effect of gravity is taken into consideration in this study.

\subsection{Minimum spatial resolution}\label{subsec:congif_spatial}
Direct Numerical Simulation (DNS) consists in resolving all scales of the flow, from the smallest relevant ones (ideally here the smallest water droplets ejected in the air or air bubbles entrained in the water) up to the large scales of the problem (here the volume of water tank that receives the initial impacting drop). Ideally, it means that the grid resolution of the air/water domain should be fine enough to capture all air/water interfaces created at all steps of the splashing process. We have carried out extensive tests to explore the effects of the minimum spatial resolution in comparison with the available experimental data. It was found that the early dynamics of liquid sheet and air entrapment in the neck region affect significantly the subsequent phenomena, such as the formation of the crown and its closure at high $Re$ and $We$ numbers. Our preliminary study has shown that a grid resolution of at least 1024 cells per drop diameter is necessary to capture the correct regime of early splashing for the present work (see Appendix \ref{appA} for details). This will need to apply a maximum level of refinement at $L_{max}=14$, which corresponds to an equivalent uniform grid of more than 4.3 trillion  $[{(2^{14})}^3]$ cells. Note that the more the maximum refinement level is increased, the more the time steps are reduced to comply with the CFL (Courant–Friedrichs–Lewy) condition, which means that the calculation CPU time is not proportional to the number of cells. Although the total number of cells reduces significantly by employing the AMR algorithm, it is still far too expensive to perform a long-term simulation in full 3D configuration at this level.

Therefore, based on the physical characteristics during the impact, we divide the simulation into three consecutive stages, namely the early-time splashing stage at $t\leqslant0.2$ ms ($S1$), the crown formation stage at $0.2<t<4$ ms ($S2$) and the bubble canopy stage at $t\geqslant4$ ms ($S3$). At stage $S1$, very thin liquid sheet emerges on the neck region and interacts strongly with the surface of drop and pool, producing numerous very fine droplets and bubbles, thus the primary objective is to capture the flow dynamics near the contact region in finer scales. Figure \ref{fig:initial}(\textit{c}) shows the mesh refinement strategy employed at the initial stage. A higher maximum level $L_{max}=14$ (3.9 $\upmu$m) is used in the vicinity of the free surface to solve the dynamics of the neck region in smaller scales, while $L_{max}=13$ (7.8 $\upmu$m) is used for the rest of the domain to capture the general dynamics. At stage $S2$, a coherent liquid sheet has been developed above the pool and it grows subsequently into the thin-walled crown. During this rapid expansion stage, a great deal of secondary droplets is sprayed incessantly from the top of the crown \citep{murphy2015splash}. The extra layer of refinement with $L_{max}=14$ is thus removed at this stage and $L_{max}=13$ is used for the entire computational domain to capture the statistics of droplets and bubbles. At stage $S3$, the droplets shed from the top of the crown are much less and generally larger, comparing with the ones from previous stages. Therefore, we restart the simulation with $L_{max}=12$ (15.6 $\upmu$m) over the whole computational domain, which allows capturing the main physical dynamics of crown and cavity till the end of the simulation (t = 48 ms). This mesh refinement strategy ensures that the simulation is doable in a full three-dimensional configuration and can be accomplished in a ``reasonable" time, considering the available computational resources for the moment. 

All the numerical results presented in the main text of this paper were performed on 1024 cores for 33.5 days, which consumes more than $8.21 \times {10}^5$ CPU-hours in total, using the computational resources on Advanced Research Computing (ARC) at Virginia Tech.

%------------------3 Validation---------------------------------------%
\section {Comparisons with the experiments}\label{sec:validation}

\subsection {Morphology}\label{subsec:valid_morph}
\begin{figure}
  \centering
  \begin{overpic}[scale=1.2]{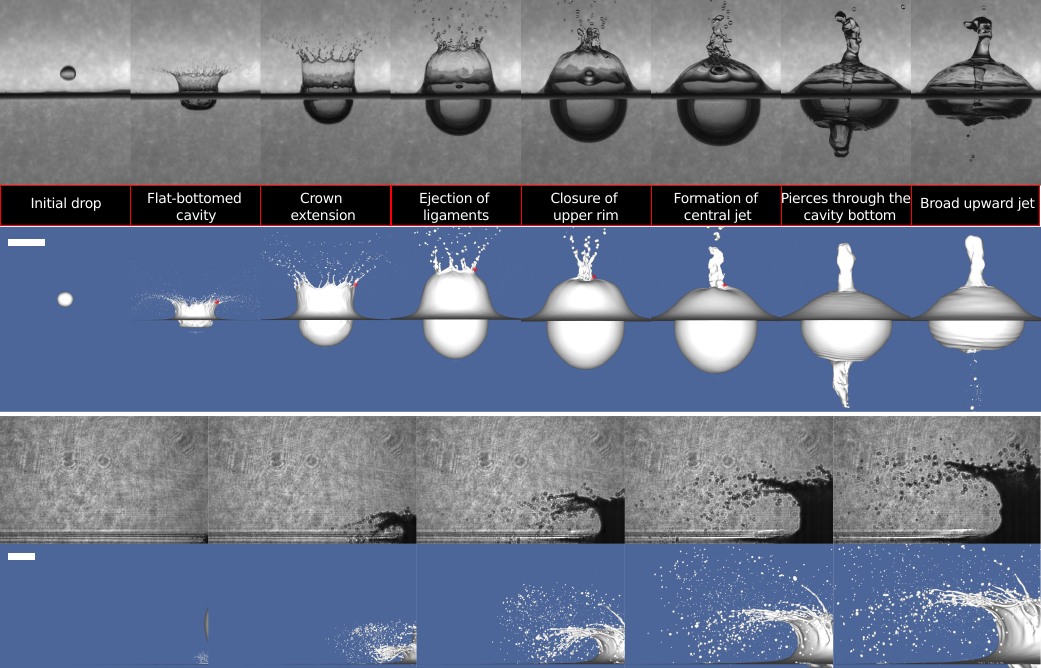} % Images in 100% size
  \put(-4,62.4){\small(\textit{a})}
  \put(-4,22.2){\small(\textit{b})}
  \end{overpic}
  \caption{Qualitative comparison between experiments and simulations. (\textit{a}) Overall dynamics of the air-water interface during 48 ms after impact. From left to right, the experiment shows -1, 1, 3, 7, 12, 18, 41 and 52 ms after impact, and the simulation shows -1, 1, 3, 7, 12, 18, 37 and 48 ms after impact. The red stars indicate the tracked positions of the upper rim of the crown. The scale bar is 10 mm long. (\textit{b}) Closeup view of the early-time splashing during 450 $\upmu$s after impact. From left to right, the experiment shows 49, 148, 246, 345 and 443 $\upmu$s after first contact, and the simulation shows 50, 150, 250, 350 and 450 $\upmu$s after first contact. The scale bar is 1 mm long. The qualitative comparisons show that the simulation successfully reproduced all the distinctive features observed in the experiments.}
  \label{fig:valid_interface}
\end{figure}

Figure \ref{fig:valid_interface}(\textit{a}) shows the global evolution of the air-water interfaces generated by high-energy drop-pool impact during the 48 ms after contact. The side-views of the numerical results (bottom) are compared with the experimental high-speed images (top) in time. The time point of contact is defined as $t=0$ ms. At $t=-1$ ms, the drop is initialised above the surface of the pool. Once the simulation starts, the drop will fall downwards to hit the liquid surface driven by the combination of the initial impact speed and the gravitational force. Directly after impact ($t=1$ ms), a cylindrical wave around a flat-bottomed disk-like cavity is produced due to the drop violent penetration and substantial liquid ligaments emanate almost horizontally from the thickened rim of the crown, spraying a large number of micro-droplets into the air. At this stage, most of the drop liquid is concentrated on the bottom of the cavity and the submergence of the drop/pool interface can be approximated as half of the initial drop impact velocity \citep{fedorchenko2004some}. In the following few milliseconds, the drop keeps expanding radially and eventually spreads into a thin layer of liquid that is distributed along the interior surface of the cavity, stretching the cavity into a typical hemispherical shape that has been widely reported in literature \citep{engel1966crater,berberovic2009drop, bisighini2010crater, ray2015regimes}, as seen at $t=3$ ms. By $t=7$ ms, the upper rim of the crown has started to proceed inwards and the orientation of the ligaments transitioned from almost horizontal to vertical, which eventually leads to the upcoming closure event on the upper part. At the time when the crown necks in ($t\approx12$ ms), flows from the sidewalls of the cylindrical wave meet along the impact axis and a large volume of air is encapsulated, generating a central liquid jet that moves spirally from the merging point ($t=18$ ms). The downward-moving jet keeps growing and eventually pierces the cavity bottom as shown at $t=37$ ms, disturbing the retraction of the cavity bottom. Finally, the rebound of the cavity produces a broad upward jet that merges with the previous central column of fluid, leaving several air bubbles inside the liquid tank, evidenced at $t=48$ ms. 

Furthermore, a close-up view of the early-time splashing near the contact region during the first 450 $\upmu$s after impact is provided and compared with the high-speed experimental holograms in figure \ref{fig:valid_interface}(\textit{b}). An immediately ruptured ``liquid sheet" appears right after contact ($t=50$ $\upmu$s), known as the ``prompt splash" \citep{deegan2007complexities}. In this process, a great number of very fine droplets are scattered in the air, which has been considered as one of the primary sources of marine aerosols and may raise potential health issues to the public under certain circumstances \citep{murphy2015splash}. More liquids are subsequently pushed out from the pool to form the liquid wall of the crown. The morphological behaviours of the early-time splashing captured by simulation are quite consistent with the experimental observations. 

These qualitative comparisons show that our simulation reproduced all the distinctive features observed in the experiments. The correct prediction of the early-time splashing, the transition of the ligaments' orientation, and the exact time when the upper rim of the crown necks in and the central spiral jet pierces the cavity bottom are especially convincing. A general good agreement to the experiments is obtained.

\subsection {Kinematics of crown and cavity}\label{subsec:valid_kine}
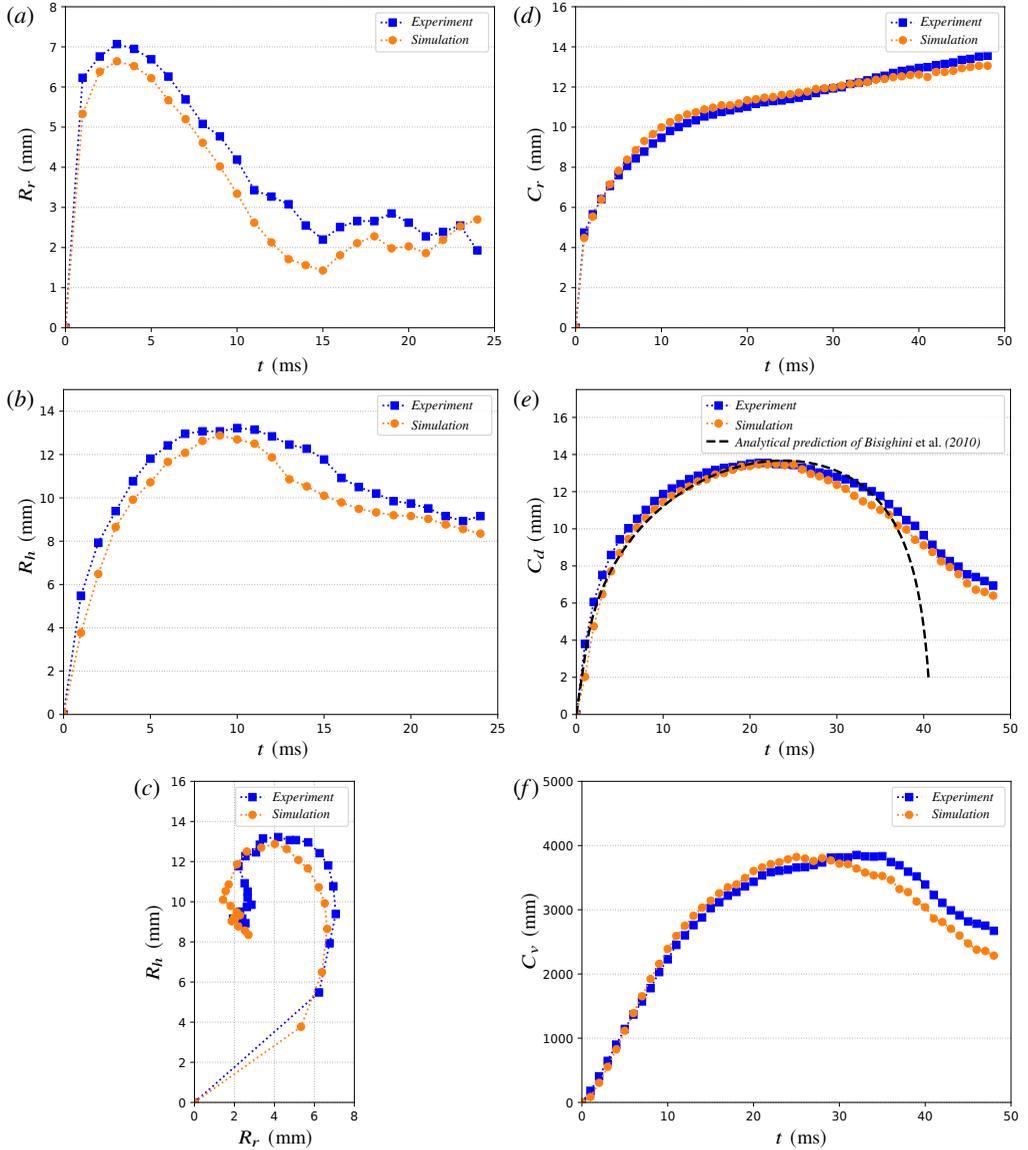
\begin{figure}
  \centering
  \begin{overpic}[width=\linewidth]{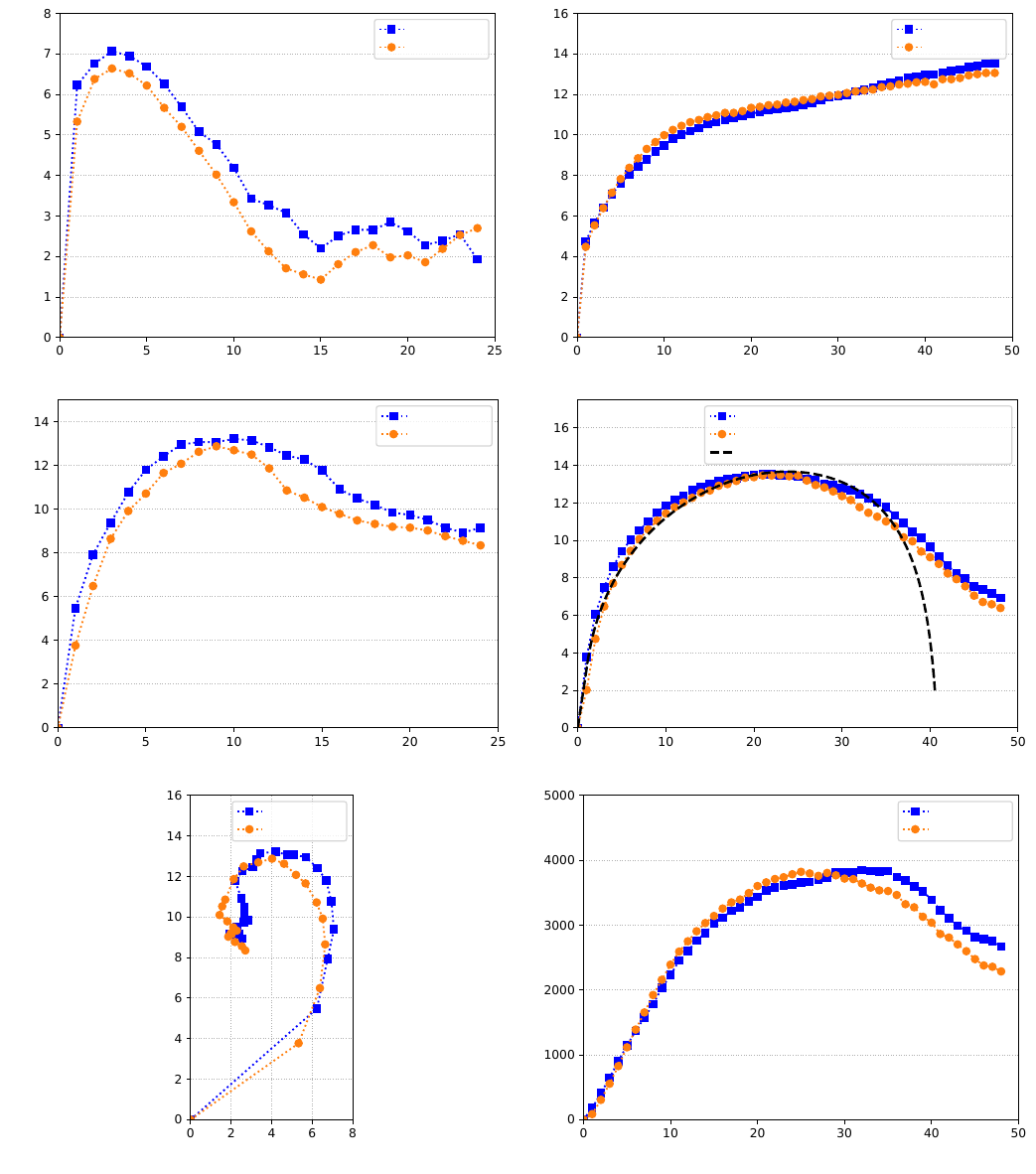} % Images in 100% size
  \put(0,97.5){\small(\textit{a})}
     \put(35,97.2){\tiny\textit{Experiment}}
     \put(35,95.7){\tiny\textit{Simulation}}
     \put(21.7,67.5){\scriptsize\textit{$t$ $\rm{(ms)}$}}
     \put(1,82){\scriptsize\rotatebox{90}{\textit{$R_{r}$ $\rm{(mm)}$}}}
  \put(0,64.5){\small(\textit{b})}
     \put(35,64.2){\tiny\textit{Experiment}}
     \put(35,62.5){\tiny\textit{Simulation}}
     \put(21.7,34.5){\scriptsize\textit{$t$ $\rm{(ms)}$}}
     \put(1,50){\scriptsize\rotatebox{90}{\textit{$R_{h}$ $\rm{(mm)}$}}}
  \put(11,31){\small(\textit{c})}
     \put(23,30.5){\tiny\textit{Experiment}}
     \put(23,29){\tiny\textit{Simulation}}
     \put(20,1){\scriptsize\textit{$R_{r}$ $\rm{(mm)}$}}
     \put(12,15){\scriptsize\rotatebox{90}{\textit{$R_{h}$ $\rm{(mm)}$}}}
  \put(43.8,97.5){\small(\textit{d})}
     \put(79,97.2){\tiny\textit{Experiment}}
     \put(79,95.7){\tiny\textit{Simulation}}
     \put(65.5,67.5){\scriptsize\textit{$t$ $\rm{(ms)}$}}
     \put(45,82){\scriptsize\rotatebox{90}{\textit{$C_{r}$ $\rm{(mm)}$}}}
  \put(43.8,64.5){\small(\textit{e})}
     \put(63,64.2){\tiny\textit{Experiment}}
     \put(63,62.5){\tiny\textit{Simulation}}
     \put(63,61){\tiny\textit{Analytical prediction of \citet{bisighini2010crater}}}
     \put(65.5,34.5){\scriptsize\textit{$t$ $\rm{(ms)}$}}
     \put(45,50){\scriptsize\rotatebox{90}{\textit{$C_{d}$ $\rm{(mm)}$}}}
  \put(43.8,31){\small(\textit{f})}
     \put(80,30.5){\tiny\textit{Experiment}}
     \put(80,29){\tiny\textit{Simulation}}
     \put(66.5,1){\scriptsize\textit{$t$ $\rm{(ms)}$}}
     \put(44.5,16){\scriptsize\rotatebox{90}{\textit{$C_{v}$ $\rm{(mm)}$}}}
  \end{overpic}
  \caption{Analysis of the quantitative data measured with respect to the initial impact centre. (\textit{a}) Evolution of the crown radius. (\textit{b}) Evolution of the crown height. (\textit{c}) Trajectory of the upper rim of the crown. (\textit{d}) Evolution of the cavity radius. (\textit{e}) Evolution of the cavity height. (\textit{f}) Evolution of the cavity volume. The black dashed line in (\textit{e}) shows the theoretical prediction of the penetration depth using the proposed model by \citet{bisighini2010crater}.}
  \label{fig:valid_kine}
\end{figure}

The kinematic behaviours of the crown and the cavity are manually tracked from experiments and simulations, and compared quantitatively in figure \ref{fig:valid_kine}. The impact centre where the drop firstly contacts the pool is used as the reference point. The time evolution of the upper rim of the crown (marked by red stars in figure \ref{fig:valid_interface}), its radial distance (figure \ref{fig:valid_kine}\textit{a}) and height (figure \ref{fig:valid_kine}\textit{b}), are measured during 24 ms after impact. The height is defined as the distance between the initial quiescent free surface of the pool and the position where ligaments are forming. The trajectory of the upper rim of the crown is then plotted in figure \ref{fig:valid_kine}(\textit{c}). For the first few milliseconds, the crown expands rapidly along the horizontal radial direction and reaches its maximum radius very soon ($t\approx3$ ms). After the maximum horizontal position, the rim starts to proceed almost immediately inwards, primarily under the effect of surface tension. Meanwhile, the leading edge of the crown rises continuously in the vertical direction and approaches its maximum height right before it necks in ($t\approx12$ ms). After the closure of the upper part, this point vibrates slightly near the impact axis along with the shrinking toroidal large air bubble.

The evolution of the submerged cavity has been thoroughly discussed in experiments, simulations and theories, and substantial attention has been particularly given to the estimation of the geometric dimensions of the cavity in previous studies \citep{engel1967initial, prosperetti1993impact, leng2001splash, berberovic2009drop, bisighini2010crater, jain2019deep}. Figure \ref{fig:valid_kine}(\textit{d}) demonstrates the temporal variation of the horizontal radius at the intersection edge between the cavity and the initial free surface. The cavity grows rapidly on the horizontal plane at the beginning due to the initial violent impingement and the outward expanding speed decreases gradually. After the closure of the crown, the increase of the cavity radius slows down and is nearly linear. Starting from $t\approx3$ ms, the horizontal radius at the bottom of the crown becomes wider than at the top, which may presumably provide an overall momentum towards the central axis, thus propelling more liquid from the receiving liquid to the sidewalls of the crater. Figure \ref{fig:valid_kine}(\textit{e}) shows the time evolution of the maximum depth of the cavity. The cavity keeps expanding in depth and reaches its maximum position 24 ms after impact, which takes almost double the time of the maximum crown position ($t\approx12$ ms). The volume of the cavity is calculated as half of an ellipsoid as used in \citet{murphy2015splash} in figure \ref{fig:valid_kine}(\textit{f}).

Figure \ref{fig:valid_kine} confirms this conclusion: both the trajectory of the edges of the upper rim of the crown and the geometric dimensions of the cavity are found in very good agreement with experimental measurements. A very reliable quantitative agreement is obtained between simulations and experiments. 

\subsection {Droplets}\label{subsec:valid_droplet}
\begin{figure}
  \centering
  \begin{overpic}[width=\linewidth]{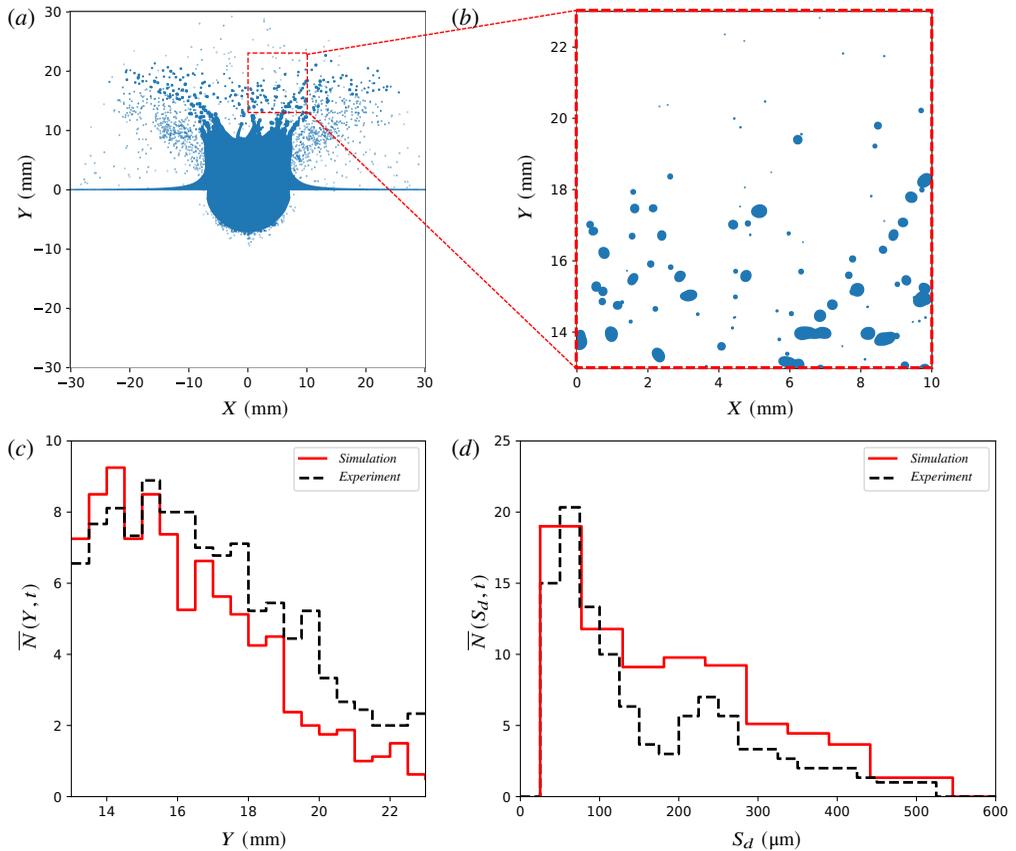} % Images in 100% size
  \put(1,81){\small(\textit{a})}
     \put(22,43){\scriptsize\textit{$X$ $\rm{(mm)}$}}
     \put(2,62){\scriptsize\rotatebox{90}{\textit{$Y$ $\rm{(mm)}$}}}
  \put(44.5,81){\small(\textit{b})}
     \put(71.5,43){\scriptsize\textit{$X$ $\rm{(mm)}$}}
     \put(51,62){\scriptsize\rotatebox{90}{\textit{$Y$ $\rm{(mm)}$}}}
  \put(1,39){\small(\textit{c})}
     \put(33.5,38.3){\tiny\textit{Simulation}}
     \put(33.5,36.5){\tiny\textit{Experiment}}
     \put(22,1){\scriptsize\textit{$Y$ $\rm{(mm)}$}}
     \put(2,20){\scriptsize\rotatebox{90}{\textit{$\overline{N}(Y,t)$}}}
  \put(44.5,39){\small(\textit{d})}
     \put(89.5,38.3){\tiny\textit{Simulation}}
     \put(89.5,36.5){\tiny\textit{Experiment}}
     \put(72,1){\scriptsize\textit{$S_d$ $\rm{(\upmu m)}$}}
     \put(46,20){\scriptsize\rotatebox{90}{\textit{$\overline{N}(S_d,t)$}}}
  \end{overpic}
  \caption{Comparisons of the droplet statistics between numerical and experimental data captured in a specific field of view. (\textit{a}) Overall schematic view of the relative position of the observation window. (\textit{b}) Closeup view of the secondary droplets in the observation window. (\textit{c}) Vertical distribution of secondary droplets. (\textit{d}) Size distribution of secondary droplets. The numerical droplet statistics presented in (\textit{c}) and (\textit{d}) are time-averaged statistics using 9 time slices over the time period $3\sim4$ ms. The experimental data are ensemble-averaged using more than 25 replications as originally presented in figure 17 and 18 in \citet{murphy2015splash}.}
  \label{fig:valid_droplet}
\end{figure}

The production of secondary droplets and their distribution induced by the process of normal (perpendicular) \citep{okawa2006production, guildenbecher2014digital, li2019characteristics, wu2020three} and oblique \citep{okawa2008effect,liu2018experimental} drop impact on a liquid surface can be found in literature, focusing mostly on the fairly ``smooth" splashing under the relatively lower range of $Re$ and $We$ numbers. The rather violent splashing behaviours observed under high-energy impacting conditions suggest that the creation of these tiny droplets might be associated with more complicated ``irregular" interfacial deformation and breakups \citep{thoraval2012karman, murphy2015splash}. Nevertheless, the understanding of the governing mechanisms and their population remain insufficient, and the available statistical data on droplet production is very limited in the literature.

Therefore, for further validating the present numerical strategies, the statistics of secondary droplets are extracted from simulations and compared with experimental data. The specific location of the observation window is shown in figure \ref{fig:valid_droplet}(\textit{a}), where a $10\times10$ $\rm{mm^2}$ square field of view is placed 13 mm above the free surface of the pool. In the experiments, the holographic frames for droplet analysis were selected at the time when the first upward-rising droplet exits the top of the observation window ($t\approx3\sim4$ ms), and the droplet statistics are then ensemble-averaged over all replicates (more than 25), as explained in \citet{murphy2015splash}. For the simulation here, the time-averaged droplet statistics are obtained using 9 time slices over the time period $3\sim4$ ms. Figure \ref{fig:valid_droplet}(\textit{c}) shows the droplet distribution in the vertical direction. Similar to the experimental observations, most of the droplets are still concentrated at a lower position, which can be clearly seen in figure \ref{fig:valid_droplet}(\textit{b}). Figure \ref{fig:valid_droplet}(\textit{d}) plots the time-averaged droplet diameter distribution. Comparing with the bimodal distribution in experiments, an inapparent second peak can be still found from the numerical results under relatively wider bins (doubling the size of bins used in \citet{murphy2015splash}). Two primary plateaux centred around 50 $\upmu$m and 225 $\upmu$m become more pronounced, which is in good agreement with the experimental data. The successful reproduction of droplet statistics in a specific observation window is highly encouraging, as it gives confidence for further comprehensive and in-depth analysis over the broader spatial domain and distinct temporal stages in the following sections.

%------------------4 Overall dynamics---------------------------------------%

\section {Overall dynamics of splashing}\label{sec:overall}

\subsection {Early-time dynamics}\label{subsec:early}
The very early dynamics that occur shortly after contact ($<100$ $\upmu$s) in the process of drop impact onto liquid surface has been widely discussed during the past thirty years \citep{weiss1999single, deegan2007complexities, liang2016review, marcotte2019ejecta}. Conventionally, for low impact velocities, an air disk is entrapped under the centre of the impacting drop by lubrication pressure and it later breaks up into chains of micro-bubbles \citep{thoroddsen2012micro, tran2013air} or contracts into several bubbles along the central line \citep{thoroddsen2003air, jian2020split} due to instability. As the contact line expands outwards radially, sheet-like liquid ejecta is sent out nearly axisymmetrically from the outer edge of the neck and possibly emits rings of tiny droplets from its rim for sufficiently large Reynolds numbers \citep{weiss1999single, thoroddsen2002ejecta, josserand2003droplet, howison2005droplet, deegan2007complexities, marcotte2019ejecta}. However, with increased impact velocity ($Re>7000$), azimuthal undulations are found to grow at the base of the ejecta and the entrapment of bubble rings takes place near the neck region \citep{thoraval2013drop, li2018early}, which thereby breaks the axisymmetry of the motions in fine scales. For even higher impact velocities, irregularly distributed splashing and incoherent liquid sheets are experimentally observed by \citet{thoroddsen2002ejecta}. Axisymmetric simulations of \citet{thoraval2012karman} have suggested that the base of the ejecta may become highly unstable under high-energy configurations, which in turn propels the ejecta base to swing to collide with drop and pool, trapping alternate air volume on both sides of the ejecta. The existence of the von K\'{a}rm\'{a}n-type vortical structure was experimentally confirmed later by \citet{castrejon2012experimental} using Shadowgraph imaging and laser-sheet visualizations. Nevertheless, the study of such complex flow structures is still very challenging for both experimental and numerical approaches as it mainly occurs in microscopic length within the time scale of several microseconds after contact.

\begin{figure}
  \centering
  \begin{overpic}[width=\linewidth]{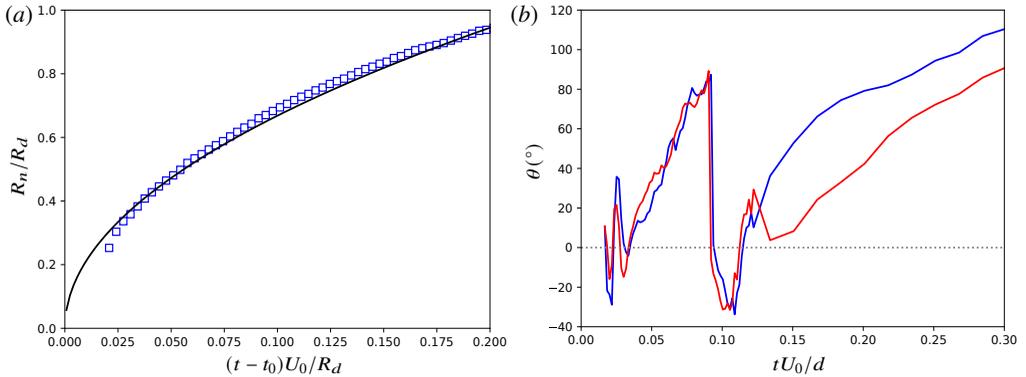} % Images in 100% size
  \put(0,35){\small(\textit{a})}
     \put(21.5,1){\scriptsize\textit{$(t-t_0)U_0/R_d$}}
     \put(0.5,18){\scriptsize\rotatebox{90}{\textit{$R_n/R_d$}}}
  \put(49.5,35){\small(\textit{b})}
     \put(75.2,1){\scriptsize\textit{$tU_0/d$}}
     \put(51,19.5){\scriptsize\rotatebox{90}{\textit{$\theta(^{\circ})$}}}
  \end{overpic}
  \caption{Early-time dynamic behaviours of the neck region. (\textit{a}) Evolution of the neck radial position $R_n$. The solid line shows the theoretical estimate using the form $R_n\sim C\sqrt{3(t-t_0)U_0/R_d}$, where $C=1.22$ and $t_0=-2.22$ are obtained by fitting the numerical measurements. (\textit{b}) Evolution of the ejecta angle $\theta$ measured from the vertical central slices as shown in figure \ref{fig:neck2D} (two sides), using the definition sketch proposed by \citet{thoraval2012karman}. The sharp decrease of the ejecta angle indicates the ``bumping" event, where the contact point of the neck suddenly changes due to the reconnection between the ejecta and the drop/pool.}
  \label{fig:neck_motion}
\end{figure}

Upon contact, most of the liquid remains unperturbed while the connection of two liquid masses propagates outwards instantaneously. Theoretically, the spreading law of $R_n\sim\sqrt{2\tau}$ can be derived from the truncated sphere approximation based only on the geometric considerations \citep{rioboo2002time, josserand2003droplet}, but it violates the continuity equation. Combining Wagner's theory \citep{wagner1932uber}, the radial motion of the neck for drop impact on a solid surface can be described using the form $R_n\sim C\sqrt{3\tau}$ \citep{riboux2014experiments, philippi2016drop, li2018early}. The dimensionless time is defined as $\tau=(t-t_0)U_0/R_d$, where $R_d=d/2$. Figure \ref{fig:neck_motion}(\textit{a}) confirms the good match between simulated neck motion with analytical prediction. The coefficient $C=1.22$ and $t_0=-2.22$ are obtained by fitting the numerical measurements. At high impacting $Re$, the initial spreading speed of the neck may reach as high as $\sim17$ times the impact velocity \citep{li2018early}, which basically requires the time resolutions of the order of $10^{-7}\sim10^{-6}$ s to observe the instantaneous motions along the contact line, possibly explaining the limited knowledge in this specific area.

\begin{figure}
  \centerline{\includegraphics[width=\linewidth]{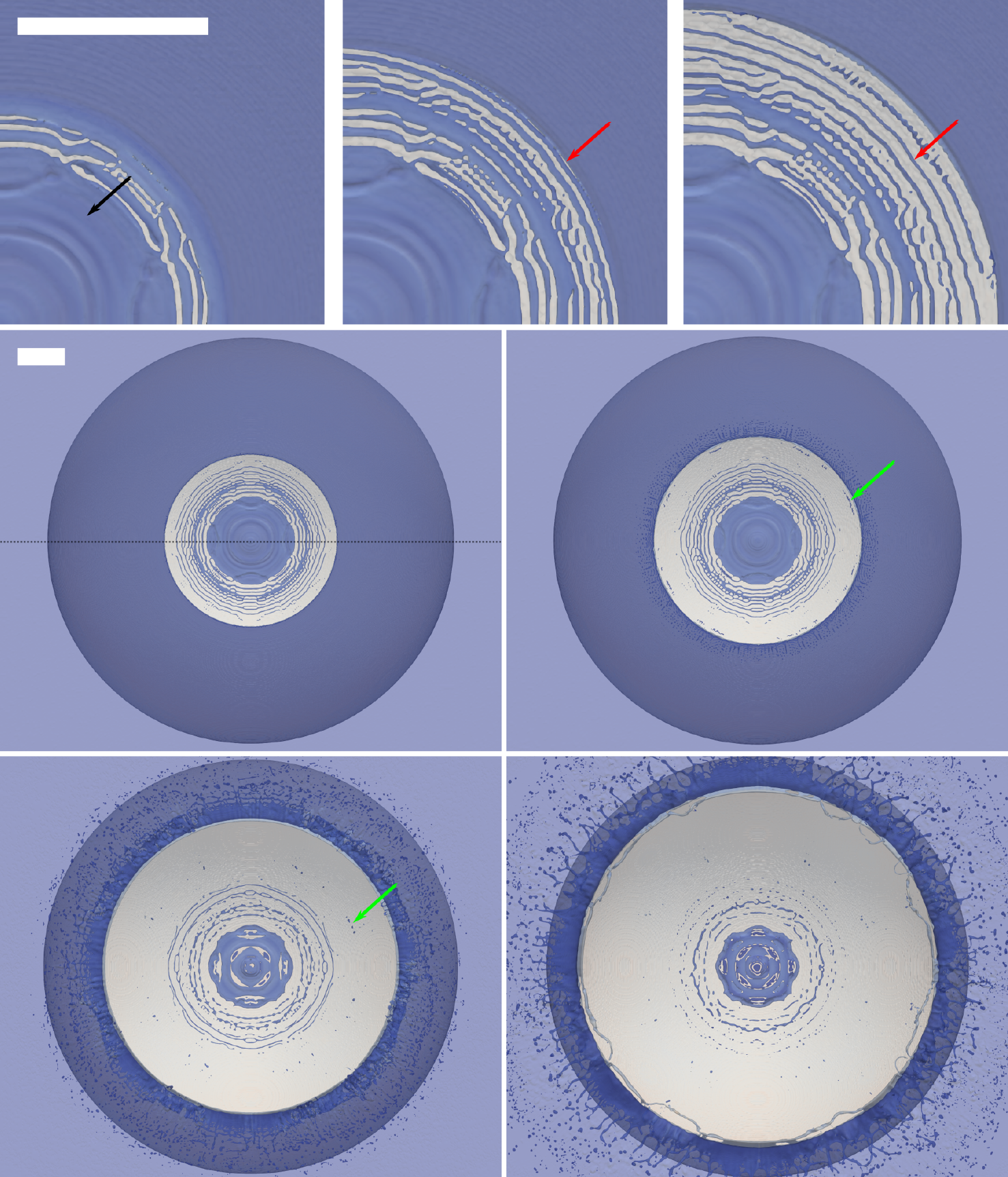}}% Images in 100% size
  \caption{Early-time dynamics near the contact region induced by high-speed drop impact onto deep liquid pool observed from the bottom view. The first three frames are shown 4, 6 and 8 $\upmu$s after the first contact, where the ``nearly axisymmetric'' bubble rings are entrapped from the neck of the connection. The black arrow points at the central air disk. The red arrows indicate the formation of a new bubble ring. The last four images show a smaller magnification 10, 15, 32 and 50 $\upmu$s after impact. The outer edge is the downward-moving drop, the inner edge is the contact line of the neck and the central irregular disc is the entrapped air sheet. Azimuthal instabilities and liquid ejecta are developed along the neck. The outer line of the neck has not reached the size of the impacting drop here. The scale bar is 500 $\upmu$m long.}
\label{fig:earlybottom}
\end{figure}

Figure \ref{fig:earlybottom} shows the simulated early-time dynamics in the neck region between drop and pool from the bottom view. Here it is focused on the stage before the outer edge of the neck overtakes the outline of the drop. The first frame and the last frame reach $25\%$ and $89\%$ of the drop size respectively. The air-water interface is coloured by the volume fraction of the passive tracer (see section \ref{subsec:tacer}) and the opacity of the interface is set to 0.5, which enables to visualize the interfacial dynamics for both the inside and outside regions of the neck. 

Immediately after the contact, a central disc of air (black arrow) is entrapped and an outer liquid edge connecting the drop and pool is formed. The contact line then expands rapidly in the radial direction, entrapping bubble rings along the drop/target boundary. In the second panel of figure \ref{fig:earlybottom}, it can be seen that a new round of air cylinder (red arrow) is entrapped at the neck and is pinched off shortly into the bulk within the time scale of $<0.4$ $\upmu$s. By $t=8$ $\upmu$s, up to 10 bubble rings are present within the target volume and most of them are entrapped axisymmetrically as concentric circles. For the next $\sim20$ $\upmu$s, these intact air rings will be stretched longitudinally while rotating, and eventually break up into a necklace of micro-bubbles due to surface-tension Rayleigh instability \citep{chandrasekhar2013hydrodynamic} as shown at $t=32$ $\upmu$s and $t=50$ $\upmu$s, which looks very similar to the experimental observations of figure 7 in \citet{thoroddsen2003air}. 
\begin{figure}
  \centering
  \begin{overpic}[scale=0.84]{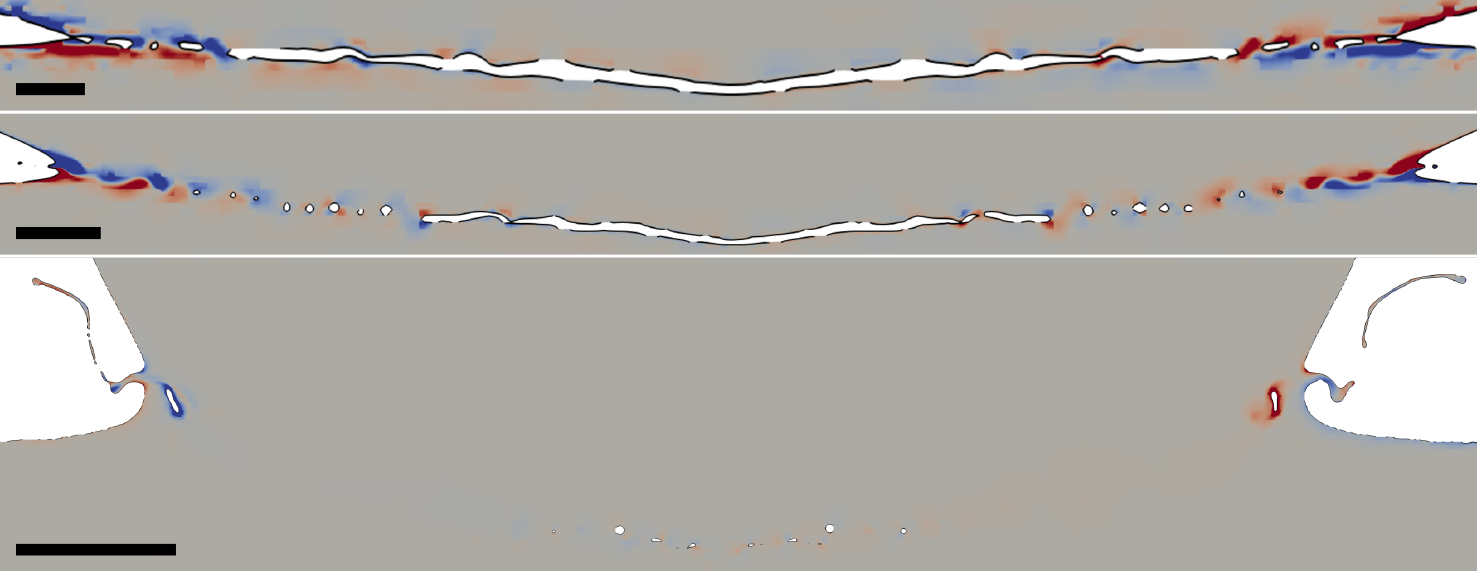} % Images in 100% size
  \put(-3.5,37){\small(\textit{a})}
  \put(-3.5,29.5){\small(\textit{b})}
  \put(-3.5,19.5){\small(\textit{c})}
  \end{overpic}
  \caption{Flow field and vorticity structure in the vicinity of the neck region on the vertical slice at $z=0$ (see dashed line at $t=10$ $\upmu$s in figure \ref{fig:earlybottom}). The red and blue colours represent counterclockwise and clockwise rotation respectively. (\textit{a}) Entrapment of air disk and bubble rings 4 $\upmu$s after the first contact. The scale bar is 50 $\upmu$m long. (\textit{b}) Formation of liquid ejecta from the neck at $t=12$ $\upmu$s. Secondary droplets are emitted from its tips. Vortex shedding of the alternate sign from the base of the ejecta generates a Von Kármán-type structure along the drop/pool boundary, with only occasional air bubbles/bubble arcs entrapment from the neck. The scale bar is 100 $\upmu$m long. (\textit{c}) Collisions between the ejecta and the downward-moving drop, so-called the ``bumping event'', which leads to the entrapment of air rings at $t=73$ $\upmu$s. The scale bar is 500 $\upmu$m long.}
  \label{fig:neck2D}
\end{figure}

Note that the neck of connection between drop and pool remains rather smooth at this time ($t<10$ $\upmu$s) and no valid liquid ejecta is observed around it, implying that the fundamental mechanism of bubble ring entrapment here differs from the jet-induced air encapsulation predicted by \citet{weiss1999single}. Figure \ref{fig:neck2D} shows the air-water interface overlaid by the vorticity field near the neck region on the vertical intersection across the drop centre. An air void is rolled up along the entire circumference of the neck and later entrapped axisymmetrically inside the liquid phase. This is reminiscent of the entrapment of toroidal bubbles numerically captured by \citet{oguz1989surface} for two drops collision, where the basic question ``Does the contact line between two approaching surfaces move outwards fast enough to prevent further contact after the initial one? " was discussed. At the moment of contact, a very small liquid bridge of radius $R_n$ with high curvature ``meniscus" connects two liquid masses and a thin outer air sheet retracts outwards rapidly driven by surface tension and local pressure gradient. A self-similar repeated reconnection of this air gap was predicted at the very early time of contact, thus enclosing a number of tiny toroidal bubbles in the liquid phase. Focusing on the viscous regime of $R_e\ll1$, $R_e=\sigma R_n/(\rho\nu^2)$, the analytical analysis of \citet{eggers1999coalescence} draws the scaling laws for the radius of the entrapped toroidal bubble $r_b\propto R_n^{3/2}$ and the width of the thin air gap connecting the bubble $r_g\propto R_n^2$. Further investigation of \citet{duchemin2003inviscid} for inviscid drop coalescence explained that the occurrence of each pinch-off event ($i_{th}$) depends only on the local dynamics of air gap width $r_g^i$, where $r_g^i=(R_n^i)^2$. The distance between the initial tip of the meniscus and the reconnection point as well as the time interval can be estimated as $r_c^i=10(R_n^i)^2$ and $t_c^i=7.6(R_n^i)^3$ respectively. The author emphasised that the reconnection ceases when $R_n>0.05$, thus $r_c^i\leqslant0.025$ and $t_c^i\leqslant0.00095$, where the time and space coordinates are scaled by $\sqrt{\rho {R_d}^3/\sigma}$ and $R_d$. The evolution of the bubble rings was however not able to be predicted by the analytical model because of their highly non-circular shape and three-dimensional rotations and stretching. Based on this theoretical model, the distances and time intervals of the neighbouring bubble rings for the present case should be $\leqslant51.25$ $\upmu$m and $\leqslant10.41$ $\upmu$s. Measurements from our simulation show that the maximum distance and time interval between bubble rings are around $\sim31$ $\upmu$m and $\sim0.5$ $\upmu$s, in the same order as the analytical estimate. It should be noted that the existence of the entrapped central air disk as well as the variation of the drop bottom curvature at the moment of impact may alter the local dynamics, but the phenomena should be qualitatively similar.

\begin{figure}
  \centerline{\includegraphics[width=\linewidth]{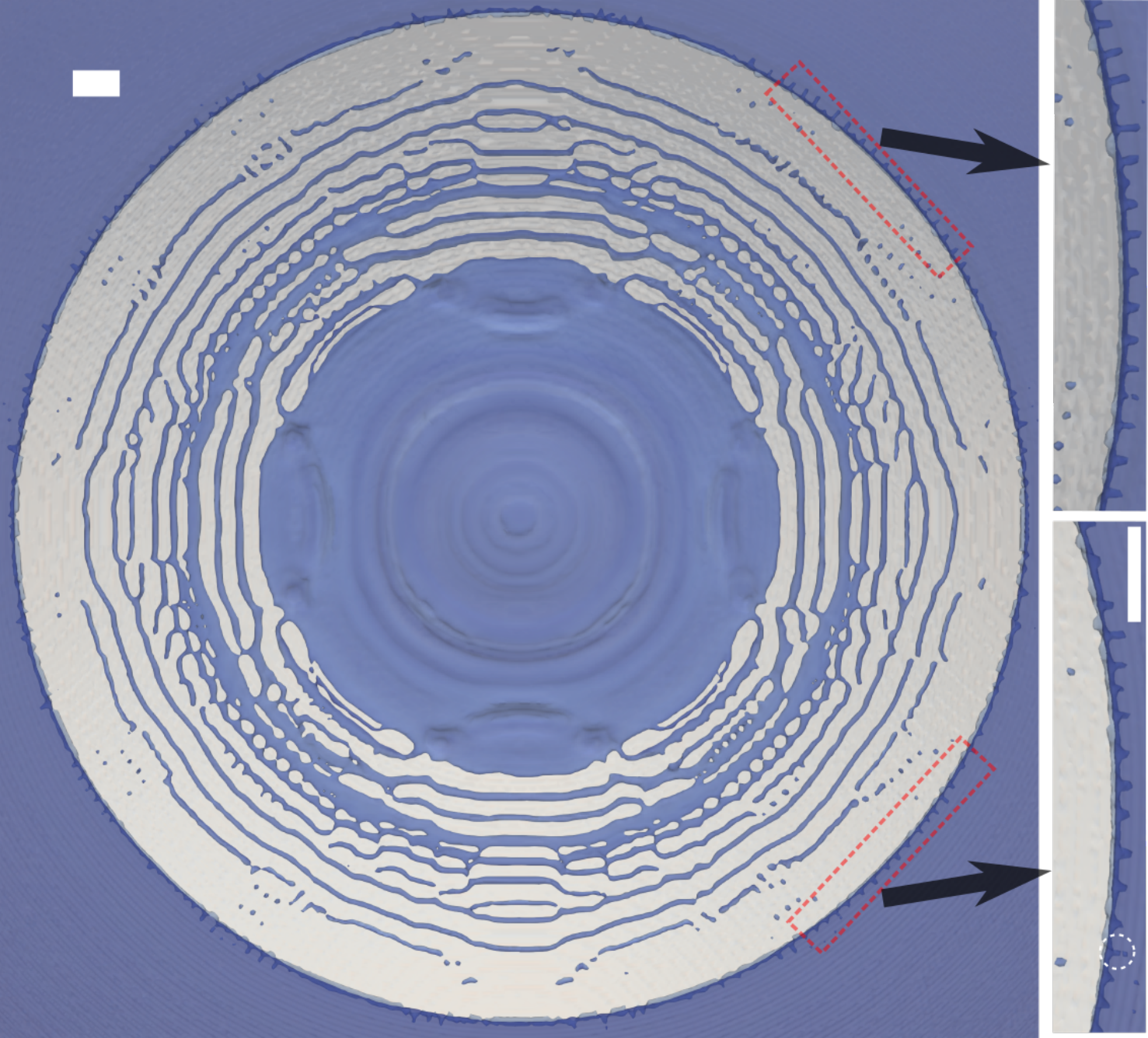}}% Images in 100% size
  \caption{Irregular azimuthal undulations on the neck region between drop and pool 10 $\upmu$s after the first contact. The central irregular plate is the entrapped air disk. The white circle indicates the early-time breakups of the ejecta fingers. The scale bar is 100 $\upmu$m long.}
\label{fig:azimathul}
\end{figure}

Starting from $t\approx10$ $\upmu$s, an azimuthal instability is observed along the neck. Here the diameter of the outer contact line reaches only $38\%$ of the drop size. Figure \ref{fig:azimathul} shows the closeup view of the azimuthal undulations at the neck of connection between drop and pool at $t=10$ $\upmu$s, where the entire periphery of the neck is visible. Irregular undulations are captured on both sides of the neck, namely the outer liquid ejecta and the inner air sheet. Liquid fingers are initiated arbitrarily from the outer side and some of them break up immediately on their tips (white dashed circle), producing the very first generation of secondary droplets (see also figure \ref{fig:neck2D}\textit{b}). At some locations where fingers are not found (or long wavelength), smooth ejecta and air sheets are present. We have measured that the characteristic wavelength and amplitude of the initial fingers are around $\sim30$ $\upmu$m and $\sim21$ $\upmu$m, which are generally longer than the thickness of the ejecta base ($\sim7$ $\upmu$m). As the neck spreads radially, alternate vortices are shed from the base of the ejecta, thus forming a similar structure of von K\'{a}rm\'{a}n-type street along the drop/target contact as shown in \ref{fig:neck2D}(\textit{b}). Meanwhile, the oscillations of the base of the ejecta pull the local air sheet on the corner of it, entrapping occasionally some isolated bubbles/bubble arcs on both/alternate sides of the base (green arrow in figure \ref{fig:earlybottom}).

After the initial small amplitude oscillations combined with the ``complete" disintegration of the ejecta jets, a coherent ejecta sheet eventually emerges (see $t=32$ $\upmu$s in figure \ref{fig:earlybottom}). This cylindrical ejecta sheet then rises radially along the contact line until it impacts the drop surface, entrapping a larger bubble ring along the neck as evidenced at $t=50$ $\upmu$s in figure \ref{fig:earlybottom} and figure \ref{fig:neck2D}(\textit{c}). Figure \ref{fig:neck_motion}(\textit{b}) plots the early-time evolution of the ejecta angle $\theta$ measured based on the definition sketch proposed in \citet{thoraval2012karman}. It can be found that $\theta$ increases almost linearly at the initial time and suddenly decreases at the moment of ``bumping", which has been also observed by \citet{thoroddsen2011droplet} and \citet{thoraval2012karman}. Once the new neck of connection is established, the fast-moving rim will stretch the ejecta sheet and tear it immediately into multiple liquid ``tori", which later break up and produce a large number of similar-sized microdroplets, as illustrated in figure \ref{fig:neck2D}(\textit{c}) and discussed later in section \ref{sec:droplet}. 

\subsection {Formation of bubble canopy}\label{subsec:BC}
\begin{figure}
  \centering
  \begin{overpic}[width=\linewidth]{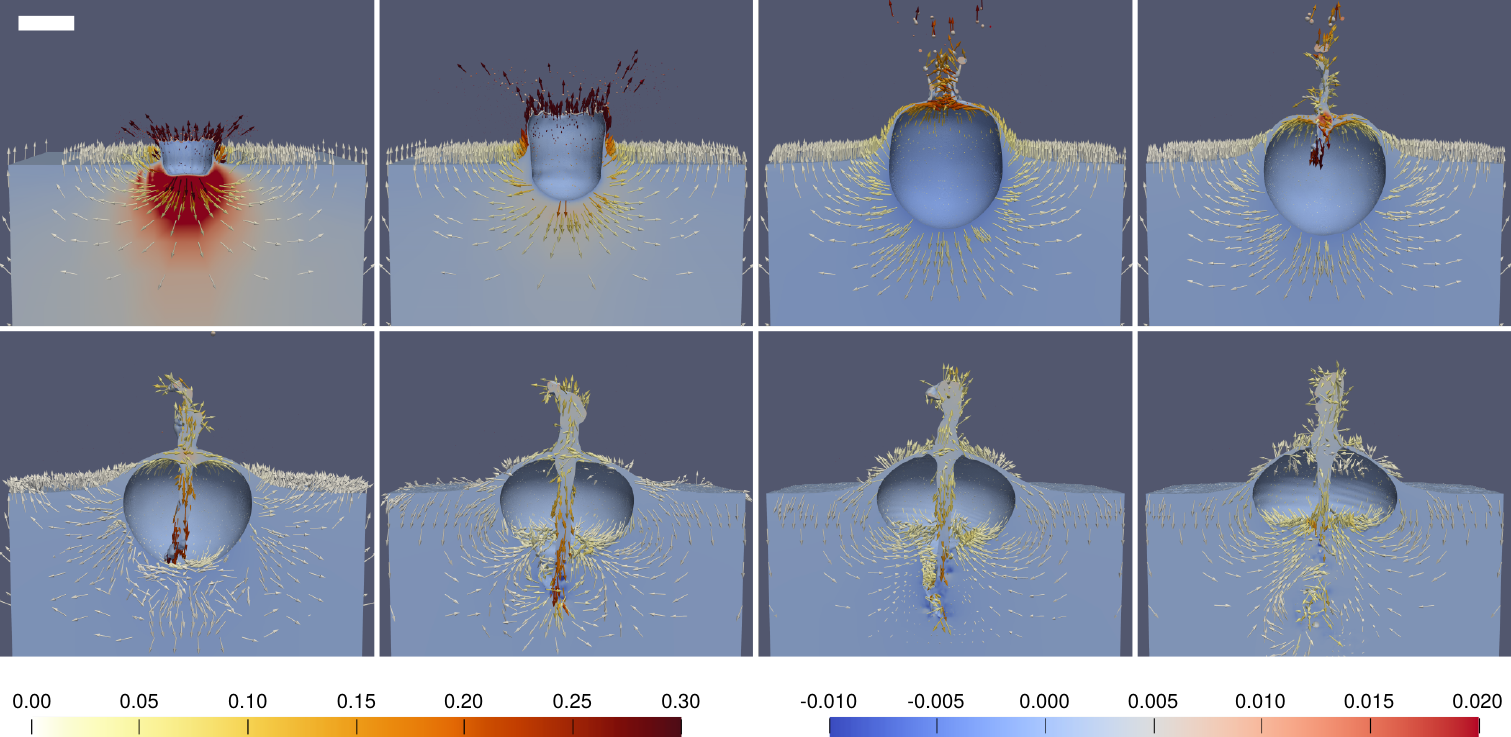} % Images in 100% size
  \put(21.4,3.4){\scriptsize\textit{$U/U_0$}}
  \put(75.2,3.4){\scriptsize\textit{$P$}}
  \end{overpic}
  \caption{Internal flows of the high-speed drop impact overlapped by velocity and pressure fields. From top to bottom and left to right, the corresponding times are 1, 3, 11, 16, 24, 32, 38 and 48 ms after impact. The velocity magnitudes are scaled by the drop impact speed $U_0$. The scale bar is 10 mm long.}
  \label{fig:Bubblecanopy}
\end{figure}

Figure \ref{fig:Bubblecanopy} shows the internal flows of the liquid phase at different stages, overlapped by velocity vector and pressure fields on the cross-section. The grid-based arrow is oriented by the velocity field and its value is represented by colour. Within the first millisecond, the drop has not yet fully sprawled out and most of the momentum is still concentrated in the impacting drop, thus generating a broad high-pressure area along the drop/pool interface. Meanwhile, an outward-expanding liquid sheet arises along the contact line as the result of the violent extrusion. As the drop moves downwards, more liquid is therefore pushed away from the pool and transported to the uprising crown. At high impact velocities, thin liquid ligaments are generated along the rim of the crown and break up on its tips by instability, producing moderate and large-scale secondary droplets (see section \ref{sec:droplet}). As discussed above in section \ref{subsec:valid_kine}, the crown reaches a maximum horizontal radial position soon after impact ($t\approx3$ ms) and then its rim thickens and bends towards the impact axis under the effect of surface tension, while it simultaneously rises in the vertical direction. Moreover, the outline of the cavity expands rapidly on the surface of the pool and overtakes the crown expansion at around $t\approx4$ ms, which also facilitates generating an inward-directed momentum on the crown top to enclose the upper part. During the period of crown expansion, the maximum velocity is reached on the top where the liquid film is thinnest. Right before the closure ($t\approx12$ ms), it is visible that the velocity on the crown rim points almost horizontally inwards. Flows from all directions on the liquid wall meet and interact at the instant of closing, generating a sharp pressure rise around the point of closure, which later protrudes upward and downward moving jets evidenced at $t=16$ ms. Meanwhile, ligaments above the dome tangle and merge, possibly shedding several large-scale droplets on the top of the dome, as shown at $t=16$ ms.

It can be found in the literature that large bubble entrapment owing to drop impact onto liquid pool occurs mainly in two types of mechanisms. Firstly, under low impact energy, a vortex-induced roll jet may be formed and grows into a thick liquid tongue, which later collapses near the pool surface to entrap the air bubble \citep{pumphrey1990entrainment}. The shape of the drop at the time of impact is crucial for such mechanism and it occurs almost exclusively for prolate-shaped drops \citep{zou2012large, wang2013we, thoraval2016vortex, deka2017regime}. Secondly, with sufficiently high impact energy, the thin-walled liquid crown rises up higher above the pool due to the violent collision and its rim bends towards the impact axis while rising vertically, enveloping the air bubble above the target pool. The formation of such complex flow structures for high-velocity drop-pool collision has been observed from time to time by experiments during the past century \citep{worthington1908study, engel1966crater, bisighini2010crater, murphy2015splash, lherm2021rayleigh}, but is probably firstly solved in 3D by high-resolution numerical simulations in the present work. Interestingly, although driven by different mechanisms, similarities are still observed between them, such as the generation of the central jets, the reconnection of the downward-moving jet as well as the bursting of the final ``floating bubble". Comparing with the low-energy counterparts, the rather intense interfacial deformation at high impact velocities surely introduces some new phenomena, which could accordingly influence the subsequent physical process like the generation of underwater noise induced by air bubble entrainment \citep{prosperetti1989underwater, prosperetti1993impact} and the additional source of airborne droplets caused by liquid film rupture \citep{resch1986marine, afeti1990distribution, resch1991film}. Further investigation and analysis could be initiated in the future to compare and contrast these seemingly similar dynamic patterns.

\begin{figure}
  \centering
  \begin{overpic}[scale=1.0]{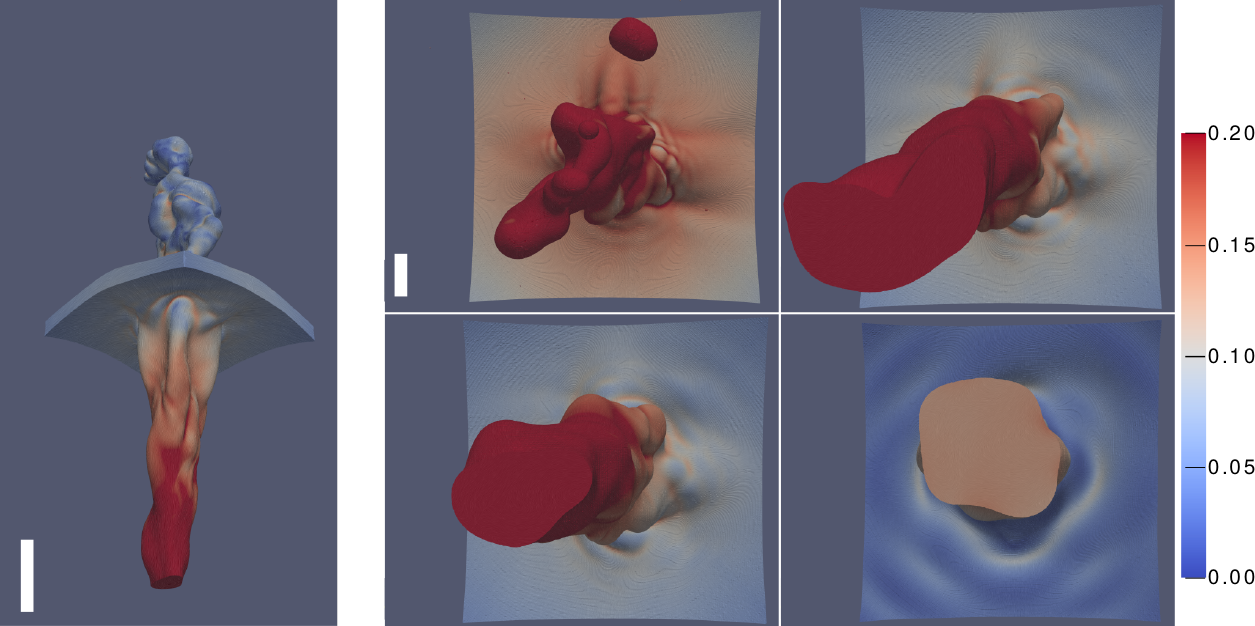} % Images in 100% size
  \put(94,42){\scriptsize\textit{$U/U_0$}}
  \put(-3,48){\small(\textit{a})}
  \put(27.2,48){\small(\textit{b})}
  \end{overpic}
  \caption{Formation of the central spiral jet inside the bubble canopy. (\textit{a}) Dynamics of the liquid jet at $t=24$ ms. The scale bar is 4 mm long. (\textit{b}) Jet motions observed from the bottom view, showing 18, 23, 24 and 38 ms after impact. The scale bar is 1 mm long. The interface is contoured by velocity field.}
  \label{fig:spiraljet}
\end{figure}

Figure \ref{fig:spiraljet} shows the formation of the central jet from the top of the dome. The jet is formed side-biased as the flows on the upper part of the crown are not perfectly axisymmetric and they do not arrive at the merging point simultaneously, which is also consistent with one of the few available experimental observations in literature \citep{bisighini2010crater, murphy2015splash, lherm2021rayleigh}. High-speed jet ejections out of a liquid interface are commonly observed in many other physical processes, such as bubble bursting \citep{boulton1993gas, thoroddsen2009spray, berny2022size}, Faraday waves \citep{hogrefe1998power, zeff2000singularity}, and cavity collapse induced by the process of solid/liquid object impact onto fluid target \citep{worthington1897v, worthington1900iv, ray2015regimes, gekle2010generation, jamali2020experimental, kim2021impact}. In general, all these jets are ejected as a consequence of a very large axial pressure gradient created at the jet base, which therefore can be further classified based on the way the large overpressure is created and the length scale at which pressure variations take place \citep{gekle2010generation}. Surface tension plays an essential role in the jet formation process.

\subsection {Cavity contraction}\label{subsec:contraction}
\begin{figure}
  \centering
  \begin{overpic}[scale=1.0]{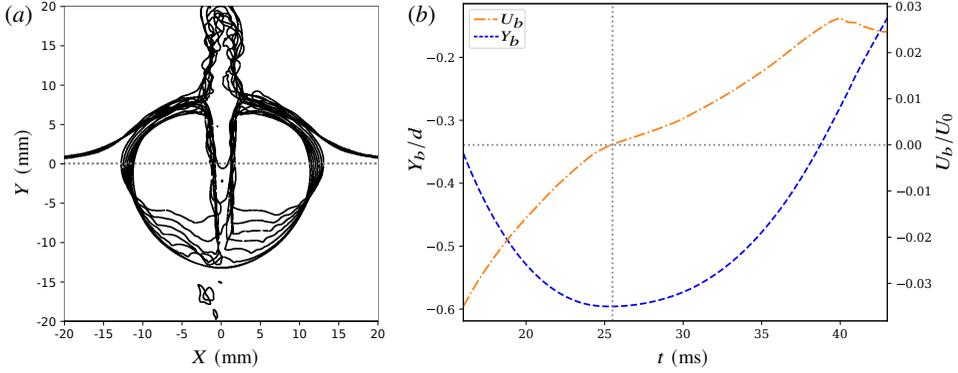} % Images in 100% size
  \put(0,36.5){\small(\textit{a})}
     \put(19.5,1){\scriptsize\textit{$X$ $\rm{(mm)}$}}
     \put(1,18.5){\scriptsize\rotatebox{90}{\textit{$Y$ $\rm{(mm)}$}}}
  \put(42,36.5){\small(\textit{b})}
     \put(51.7,36.3){\tiny\textit{$U_b$}}
     \put(51.7,34.5){\tiny\textit{$Y_b$}}
     \put(68,1){\scriptsize\textit{$t$ $\rm{(ms)}$}}
     \put(42,21){\scriptsize\rotatebox{90}{\textit{$Y_b/d$}}}
     \put(97,21){\scriptsize\rotatebox{90}{\textit{$U_b/U_0$}}}
  \end{overpic}
  \caption{Kinematic behaviours of the entrapped large bubble. (\textit{a}) Successive positions of the vertical slices for the entrapped large bubble. The time interval between each curve is 4 ms. (\textit{b}) Time evolution of the vertical centroid position (left) and the vertical speed (right) of the entrapped large bubble. The bubble sinks at the expansion stage and then starts to shallow from its bottom due to the concentric axial pressure, which eventually leads to a floating air bubble above the pool surface.}
  \label{fig:BC_position}
\end{figure}

Now we focus on the contraction of the cavity. For drop impact at low and moderate velocities, the liquid crown collapses and generates capillary waves that move radially towards the cavity bottom along the interior of the crater after it reaches the maximum expansion. The capillary waves then meet concentrically at the bottom of the cavity, forming a classic upward Worthington jet that may break up on its tips \citep{ray2015regimes}. At higher impact velocities, qualitatively different phenomena are observed. As shown in figure \ref{fig:Bubblecanopy}, during the cavity expansion stage, the direction of the velocity field is outward and upward around the enlarging cavity and the cavity continues to expand even after the closure of the upper part. The cavity depth reaches its maximum value at $t\approx24$ ms as the cavity radius extends continuously along its edge (see also figure \ref{fig:valid_kine}\textit{e}), which is very different from the measurements for an even more energetic case impacting at $\sim2U_0$ in \citet{engel1966crater} where the crown and cavity arrive at their maximum positions at about the same time. At the moment that the penetration depth reaches its maximum value, the flow around the cavity bottom should have reached its minimum value of essentially zero and shortly be redistributed by inertia force, as seen at $t=24$ ms in figure \ref{fig:Bubblecanopy}. At approximately the same time ($t\approx25$ ms), the downward spiral jet arrives at the cavity bottom and penetrates deeply into the bulk, creating a subcavity that moves also spirally, which may potentially accelerate the re-establishment of the velocity field. By $t\approx32$ ms, it can be observed that the velocity field has been completely reversed in the pool and a new circulation has been established. The flow around the cavity in the pool has transitioned from outward-upward expansion to inward-downward contraction. Such a change of the flow direction would result in a pressure buildup around the lower part of the cavity, thus pushing the cavity to shallow. Air bubbles are entrapped in the bulk in this process as can be seen at $t=32$, 38 and 48 ms in figure \ref{fig:Bubblecanopy} and figure \ref{fig:BC_position}(\textit{a}). In the last frame of figure \ref{fig:Bubblecanopy}, an upward jet eventually rises from the cavity floor and merges with the previous downward jet.

Figure \ref{fig:BC_position}(\textit{a}) draws the successive shapes of the entrapped large bubble and (\textit{b}) plots the temporal motion of the bubble centroid in the vertical direction, showing the shallowing steps of the cavity. The bubble expands to a maximum position and then contracts from its bottom, eventually generating a toroidal air bubble floating at the top of the pool.

As for the final collapsing stage, the central column thickens and merges with the outer bubble wall, creating a horseshoe-shaped bubble that transforms later into a hemisphere. For the next long period of time (more than 300 ms), this toroidal bubble will be stretched and thinned under the action of surface tension and eventually ruptures due to instabilities. Subsequently, the film cap recedes along the periphery of the ``ruptured hole" from one side, scattering fine scale liquid droplets from the ``receding rim" to the air, as shown in the experiments by \citet{murphy2015splash} (figure 3). The production of such tiny droplets from the receding films has been also studied by \citet{lhuissier2012bursting} and \citet{dasouqi2021bursting}.

\subsection {Energy budget}\label{subsec:energy} 
As reviewed by \citet{veron2015ocean}, one of the primary motivations for studying the liquid droplets ejection is to estimate the exchange of momentum, heat and eventually mass transfer through gas-liquid interface induced by the interfacial processes such as large scale breaking wave event \citep{mostert2022high}, small scale bubble bursting \citep{berny2022size} and drop impact \citep{liang2016review} etc. In the present work, the kinetic energy $E_k$, gravitational potential energy $E_g$ and surface potential energy $E_s$ of the liquid pool are calculated as follows:
\begin{equation}
  E_k=\frac{1}{2}\displaystyle \int_{V_p}\rho\|\boldsymbol{U}\|^2dV_p   
  \label{eq:energy_ke}
\end{equation}
\begin{equation}
  E_g= \displaystyle \int_{V_p}\rho \boldsymbol{g}ydV_p-E_{g0}  
  \label{eq:energy_po}
\end{equation}
\begin{equation}
  E_s= \displaystyle \int_{S_p}\sigma dS_p-E_{s0}  
  \label{eq:energy_sur}
\end{equation}
The integrals are computed over the volume ($V_p$) and surface ($S_p$) of the largest liquid continuum in the domain, which means that the small droplets detached from the pool and the gas phase are not included here. The time point when the first contact occurs between drop and pool is defined as $t=0$ ms. $E_{g0}$ and $E_{s0}$ are the gravitational and surface potential energies of the liquid phase at $t=0$ ms. In addition, the kinetic, gravitational and surface potential energies of the ejected droplets are integrated over each droplet ($V_{p}^i$ and $S_{p}^i$, respectively) using the same equations like \ref{eq:energy_ke}, \ref{eq:energy_po} and \ref{eq:energy_sur}, and the sum of energy carried by small droplets is represented as $E_d$. The total energy of the liquid phase is therefore calculated as $E_T=E_k+E_g+E_s+E_d$. $E_0$ is the initial kinetic energy introduced by the impacting drop.

\begin{figure}
  \centering
  \begin{overpic}[scale=1.3]{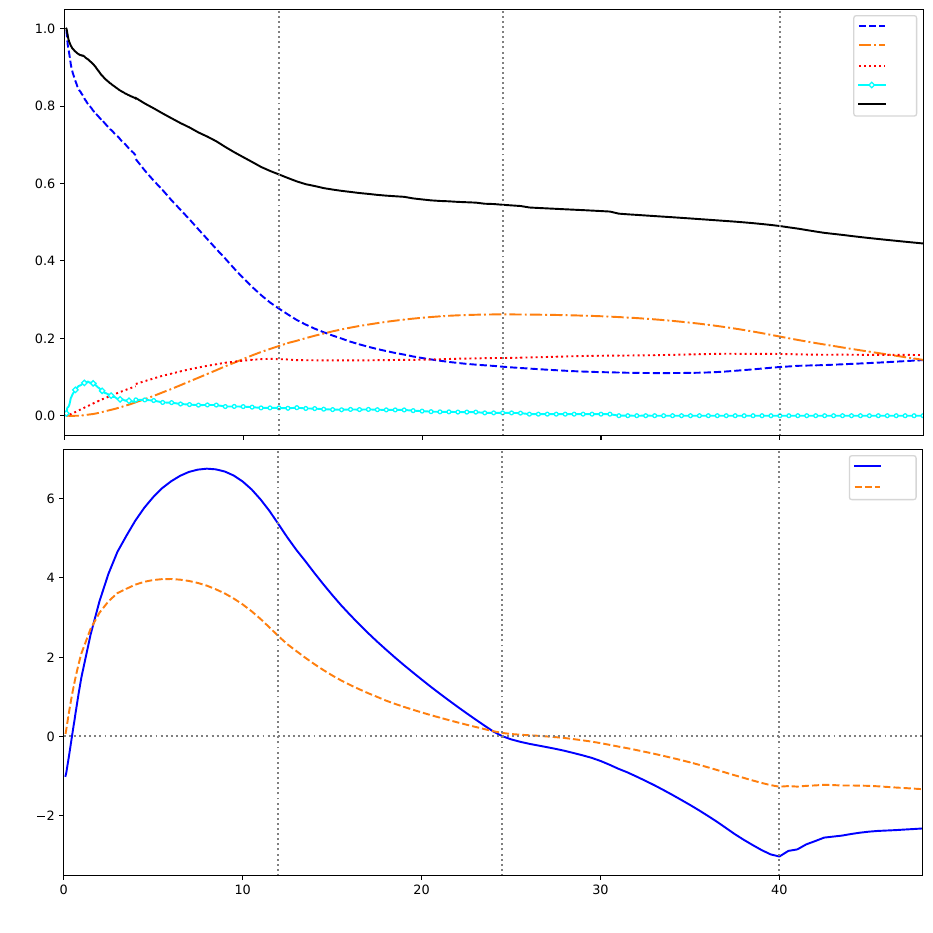} % Images in 100% size
  \put(0,97.5){\small(\textit{a})}
     \put(95.5,96.7){\tiny\textit{$E_k$}}
     \put(95.5,94.7){\tiny\textit{$E_g$}}
     \put(95.5,92.4){\tiny\textit{$E_s$}}
     \put(95.5,90.4){\tiny\textit{$E_d$}}
     \put(95.5,88.3){\tiny\textit{$E_T$}}
     \put(1,74){\scriptsize\rotatebox{90}{\textit{$E/E_0$}}}
  \put(0,50){\small(\textit{b})}
     \put(95,49.5){\tiny\textit{$P_y$}}
     \put(95,47.3){\tiny\textit{$P_{xx}$}}
     \put(51,1){\scriptsize\textit{$t$ $\rm{(ms)}$}}
     \put(1,27.5){\scriptsize\rotatebox{90}{\textit{$P/P_0$}}}
  \end{overpic}
  \caption{Energy budget for the process of high-speed drop impact onto deep liquid pool. (\textit{a}) Temporal evolution of the energy budget in liquid phase. The energies are normalised by the initial kinetic energy of the impacting drop. (\textit{b}) Temporal evolution of the momentum components in the pool (largest liquid continuum), normalised by the initial impact momentum. From left to right, the vertical dotted lines indicate the time point for the closure of the upper rim of the crown ($t\approx12$ $\upmu$s), the connection of the downward spiral jet to the cavity bottom ($t\approx24$ $\upmu$s) and the formation of the central upward jet from the cavity floor ($t\approx40$ $\upmu$s).}
  \label{fig:main_energy}
\end{figure}
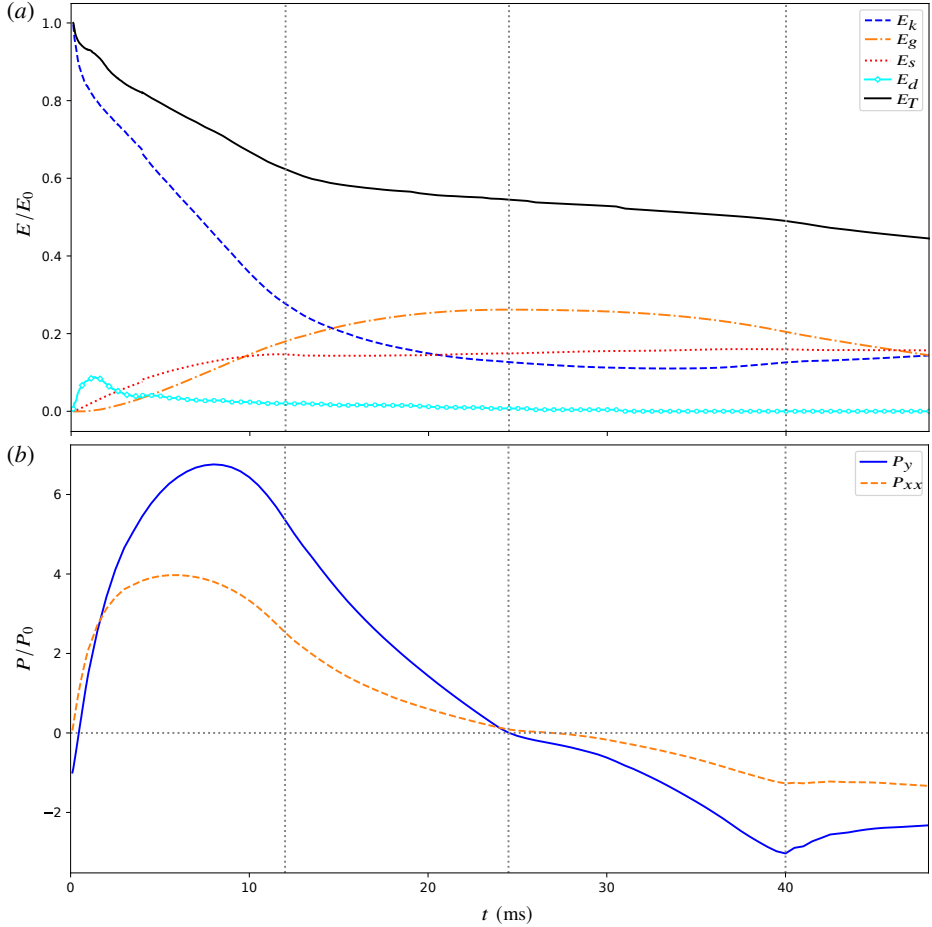

Figure \ref{fig:main_energy}(\textit{a}) plots the temporal evolution of the energy aspects throughout the simulation. The calculated energies in the liquid phase are normalized by the initial kinetic energy of the impacting drop $E_0$. Initially, at the time the drop hits the pool, the system's energy is mainly composed of the kinetic energy of the impacting drop. As the drop moves downwards, a sharp decrease of the kinetic energy is observed at the very early time of impact, which can be associated with the immediate ``prompt splash" of the liquid fragmentation ($E_d$) and the energy dissipation due to strong vortical entanglement near the neck region. The gravitational potential energy remains insignificant at this stage since the splashing appears mostly near the pool surface. Subsequently, a period of rapid expansion of crown and cavity takes place (see also figure \ref{fig:valid_interface} and \ref{fig:Bubblecanopy}), which is reflected by the almost linear decrease of the kinetic energy and noticeable increase of the gravitational and surface potential energies. The droplet energy $E_d$ reaches its maximum value at approximately the same time when the droplet count reaches its peak during the sustained droplet shedding stage (see also section \ref{sec:droplet}). Afterwards, the number of droplets reduces continuously and $E_d$ becomes insignificant. Here, the maximum total droplet energy accounts for about $8\%$ of the initial kinetic energy, which is quite close to the theoretical premises by \citet{engel1967initial} that around $5\%$ of the initial impact energy is carried away by secondary droplets. At the instant of closing (around $t\approx12$ ms), more than $60\%$ of the initial energy is still present in the liquid phase. Once the crown necks in, changes of the slope are observed in both $E_k$ and $E_s$, where the decreasing speed of $E_k$ slows down gradually and $E_s$ reaches a plateau. $E_g$ reaches a maximum value at approximately the same time when the cavity expands to the maximum position ($t\approx24$ ms) and it decreases hereafter due to the inverse flow around the bubble canopy (see figure \ref{fig:Bubblecanopy}). Meanwhile, the kinetic energy in the pool keeps almost constant during the initial retracting stage ($t\approx24\sim40$ ms) and increases slightly later due to the protrusion of the broad upward jet from the cavity floor. The total energy $E_T$ decreases monotonically throughout the entire process and nearly half of the initial energy is eventually converted to other forms, including viscous energy dissipation, airflow/bubble energy as well as the energy contained by the removed droplets. 

To better understand the kinematic behaviours of the liquid bulk, we also extract the momentum components of the largest liquid continuum (except small droplets) as shown in figure \ref{fig:main_energy}(\textit{b}). The total vertical momentum $P_y$ can be used to describe the overall vertical motion of the liquid bulk and the total momentum on the right half of the bulk along the $x$-axis $P_{xx}$ can be used to describe its horizontal expansion, which are both normalised by the initial impact momentum of the drop $P_0$. For $t<25$ ms, a general outward-upward motion of the liquid bulk is shown in the graph, where the maximum horizontal and vertical expansion speeds are found at $t\approx6$ ms and $t\approx8$ ms respectively. Starting from $t\approx25$ ms, the bulk sinks downwards and the liquid starts to flow concentrically towards the impact axis, squeezing the large air bubble to float upwards as demonstrated in figure \ref{fig:BC_position}(\textit{b}). A turning point of $P_y$ is found at $t\approx40$ ms, which indicates the onset of the broad upward jet from the cavity floor. 

\subsection {Transportation of drop liquid}\label{subsec:tacer}
\begin{figure}
  \centerline{\includegraphics[width=\linewidth]{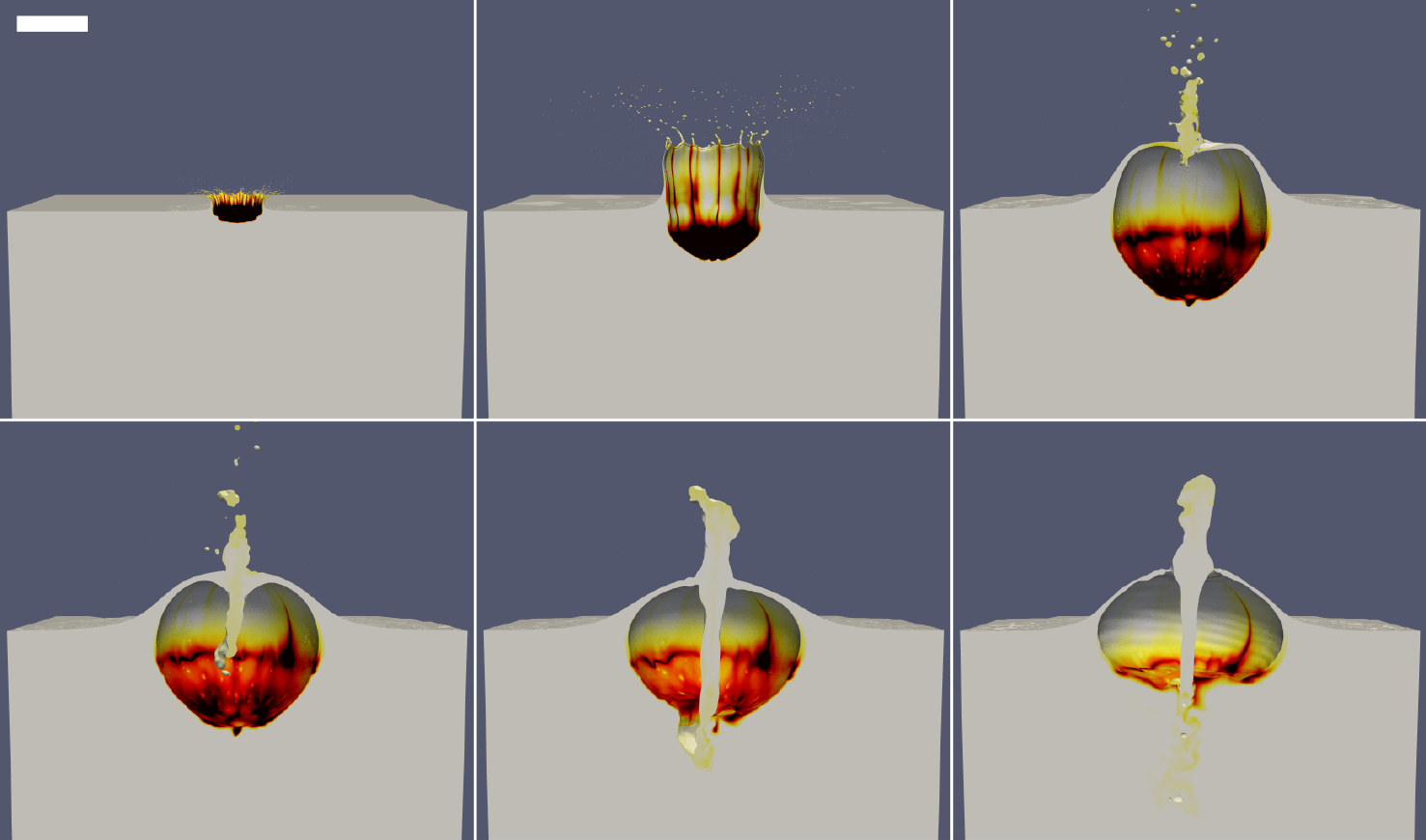}}% Images in 100% size
  \caption{Transportation of the passive tracer for drop liquid. From left to right and top to bottom, the corresponding times are 0.5, 3, 14, 20, 32 and 48 ms after impact. The scale bar is 10 mm long. Azimuthal destabilization is captured at the edge of the drop liquid, which therefore produces secondary droplets from its tips. At the later stage of impact, the thin drop film is penetrated and mixed by the downward-moving jet, producing the drop liquid plume deep inside the target pool.}
\label{fig:tracer}
\end{figure}

In the simulation, a passive tracer field $f_p$ is initially added to the impacting drop for the purpose of following the trace of the drop liquid. Experimentally speaking, this can be achieved by adding colours/dye to the liquids \citep{engel1966crater,thoroddsen2002ejecta, bisighini2011high}. The tracer field is then advected by the following equation:
\begin{equation}
  \frac{\p f_p}{\p t}+\boldsymbol{U_f}\bcdot\nabla f_p=0
  \label{eq:tracer_advection}
\end{equation}

Figure \ref{fig:tracer} shows the distribution of drop liquid at different stages of impact. Shortly after contact, the main part of the drop sits on the ``pool cavity" and the radially spreading thin film extends from its edges ($t=0.5$ ms). For the next few milliseconds, the drop deforms and expands rapidly into a thin liquid layer coated along the interior surface of the pool. Meanwhile, liquid threads are emitted from the rim of the drop liquid and then break up into small droplets on its tips ($t=3$ ms), which is comparable with the formation of azimuthal destabilization on the retracting flattened drop edge at relatively low impact energy reported in \citet{lhuissier2013drop}. By the time of closing, it can be observed that these ``drop threads" meet at the closure point and are later transported backwards to the bulk by the downward-moving jet. At the receding stage, the thin drop film starts to propagate towards the cavity bottom due to the surface-tension capillary waves and is later transported deep into the pool when the central spiral jet impinges the cavity bottom, seen at $t=32$ ms and 48 ms in figure \ref{fig:tracer}.

\begin{figure}
  \centering
  \begin{overpic}[width=\linewidth]{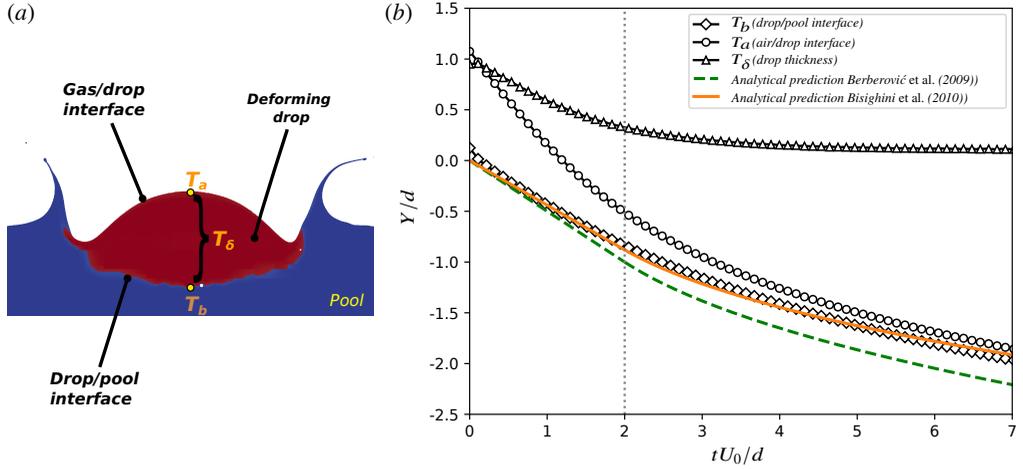} % Images in 100% size
  \put(0,45){\small(\textit{a})}
  \put(37,45){\small(\textit{b})}
     \put(71,44.3){\tiny\textit{$T_b$(drop/pool interface)}}
     \put(71,42.4){\tiny\textit{$T_a$(air/drop interface)}}
     \put(71,40.7){\tiny\textit{$T_{\delta}$(drop thickness)}}
     \put(71,38.8){\tiny\textit{Analytical prediction \citet{berberovic2009drop})}}
     \put(71,37){\tiny\textit{Analytical prediction \citet{bisighini2010crater})}}
     \put(69,2){\scriptsize\textit{$tU_0/d$}}
     \put(38.5,25){\scriptsize\rotatebox{90}{\textit{$Y/d$}}}
  \end{overpic}
  \caption{(\textit{a}) Sketch of the drop penetration. Boundaries between different fluid components are differentiated by isosurface $f_p=0.5$. The positions of the upper point $T_a$, lower point $T_b$ and the thickness of the drop tracer $T_\delta$ along the vertical axis of symmetry are tracked. (\textit{b}) Time variations of $T_a$, $T_b$ and $T_\delta$ along the axial direction. The dashed line shows the asymptotic solution proposed by \citet{berberovic2009drop}. The solid line shows the theoretical estimation of the penetration depth proposed by \citet{bisighini2010crater}. The dimensionless time $tU_0/d=2$ is indicated by vertical dotted line.}
  \label{fig:tracer_drop/poolface}
\end{figure}

Different from the gas/liquid interface that can be recorded directly by the camera, the ``virtual" drop/pool interface is usually invisible if the fluids in the drop and the receiving pool are the same, but the estimation of the kinematic behaviours of this layer is vital for theoretical studies of crater evolution \citep{bisighini2011high}. As indicated in figure \ref{fig:tracer_drop/poolface}(\textit{a}), the boundaries between different components can be easily differentiated here by the isosurface $f_p=0.5$. Figure \ref{fig:tracer_drop/poolface}(\textit{b}) plots the temporal variation for the positions of the upper point $T_a$, lower point $T_b$ as well as the thickness of the drop tracer $T_\delta$ along the impact axis. As expected, the upper point $T_a$ moves rapidly adjoining the air while the penetration speed of the lower point $T_b$ is greatly decelerated by the reacting flows from the target liquid, thus causing the decrease in drop thickness. It can be seen that the drop deforms significantly and its thickness decreases noticeably during an initial dimensionless time period $tU_0/d\leqslant2$. For the later stages, $T_\delta$ reaches a plateau and will not experience much difference throughout the expansion stage. Slight fluctuations of $T_\delta$ can be anticipated while the drop recedes into the cavity bottom at the contraction stage.

Following previous investigations in literature, a critical dimensionless time $tU_0/d\approx2$ is usually used to subdivide the evolution of drop impact into two phases \citep{fedorchenko2004some}. During the first phase ($tU_0/d\leqslant2$), the drop deforms and extends above the bulk cavity, and the interfaces of air/drop and drop/pool are clear-cut. The penetration speed of the drop/target interface during this time can be approximated as half of the impact speed $U_p\approx U_0/2$, which is well known from the penetration mechanics \citep{birkhoff1948explosives, yarin1995penetration} and has been previously applied for the analytical study of drop impact \citep{fedorchenko2004some, berberovic2009drop}. At times $tU_0/d>2$, the flow effects in the thin drop layer are negligibly small and the cavity expansion thus can be approximated as the shape of the drop/pool interface. \citet{berberovic2009drop} developed a theoretical approach to estimate the penetration depth of drop impact $T_d$ based only on the linear momentum balance of the liquid around the cavity and gave an asymptotic solution as $T_d=2^{-4/5}(5t-6)^{2/5}$ for $tU_0/d>2$ at high $Fr$, $We$ and $Re$ numbers. The predicted results using the above asymptotic formula are shown in figure \ref{fig:tracer_drop/poolface}(\textit{b}) (dashed line) and a fairly good agreement is found with the simulated results. Since the effects of surface tension, viscosity and gravity are neglected in this theory, it can not predict accurately the later stages of impact where the deformation of the shape of the cavity becomes significant due to gravity and capillary waves. \citet{bisighini2010crater} proposed a theoretical model based on the potential flow theory that accounted for the effects of inertia, gravity, viscosity and surface tension for sufficiently high Reynolds and Weber numbers, using the combination of the sphere expansion and its translation along impact axis. A system of ordinary differential equations is obtained and numerically solved using initial conditions:
\begin{equation}
  \ddot{\alpha}=-\frac{3}{2}\frac{\dot{\alpha}^2}{\alpha}-\frac{2}{\alpha^2We}-\frac{1}{Fr}\frac{\zeta}{\alpha}+\frac{7}{4}\frac{\dot{\zeta}^2}{\alpha}-\frac{4\dot{\alpha}}{\alpha^2Re}
  \label{eq:Bisignini_alpha}
\end{equation}
\begin{equation}
  \ddot{\zeta}=-3\frac{\dot{\alpha}\dot{\zeta}}{\alpha}-\frac{9}{2}\frac{\dot{\zeta}^2}{\alpha}-\frac{2}{Fr}-\frac{12\dot{\zeta}}{\alpha^2Re}
  \label{eq:Bisignini_zata}
\end{equation}
Where $\alpha$ and $\zeta$ donate the dimensionless crater radius and axial coordinate of the centre of the sphere and the dimensionless penetration depth is expressed as $\alpha+\zeta$. As explained by \citet{bisighini2010crater}, the initial conditions can be obtained from the initial phase ($tU_0/d\leqslant2$) using the forms: $\dot{\alpha}\approx0.17$, $\alpha\approx\alpha+0.17\tau$, $\dot{\zeta}\approx0.27$,  $\zeta\approx-\alpha_0+0.17\tau$, and the dimensionless width of the cavity can be estimated using the geometrical conditions $W=2\sqrt{\alpha^2-\zeta^2}\approx2\sqrt{(\alpha_0+0.17\tau)^2-(0.27\tau-\alpha_0)^2}$. By fitting the simulated bulk cavity width using the least-mean-square method, the constant is obtained as $\alpha_0=0.79$ in the present case, thus the initial conditions for equations \ref{eq:Bisignini_alpha} and \ref{eq:Bisignini_zata} are
$\alpha(2)=1.13$, $\dot{\alpha}(2)=0.17$, $\zeta(2)=-0.25$ and $\dot{\zeta}(2)=0.27$. As shown in the solid line in figure \ref{fig:tracer_drop/poolface}(\textit{b}), the temporal variation of the predicted depth of drop/pool interface agrees very well with our numerical results during the expansion stage of impact. As for the retraction stage, discrepancies between the theoretical prediction and the simulation/experiment become more pronounced as demonstrated in figure \ref{fig:valid_kine}(\textit{e}), which can be explained as the fact that the shape of the cavity does not follow the spherical expansion anymore due to the influence of the central spiral jet and the propagation of capillary waves, the model therefore is invalid for the retraction phase.

%------------------5 Droplets---------------------------------------%

\section {Airborne Droplets}\label{sec:droplet}

\subsection {Source of secondary droplet}\label{subsec:dropproduction}
According to the investigation of \citet{deegan2007complexities} in the parameter range $Re \leq 5000$ and $We \leq 1400$, at least three sources of secondary droplets can be identified during the impact: (\romannumeral1) prompt instability of the ejecta sheet occurring immediately after contact, which produces very small droplets, (\romannumeral2) rim instability of the ejecta sheet that produces medium-sized droplets and (\romannumeral3) rim instability of the crown that produces large droplets from liquid jets. These different mechanisms are typically interdependent and the earlier ones influence the later ones, which therefore further complicates the characterization of these tiny airborne droplets. 

\begin{figure}
  \centering
  \begin{overpic}[scale=0.75]{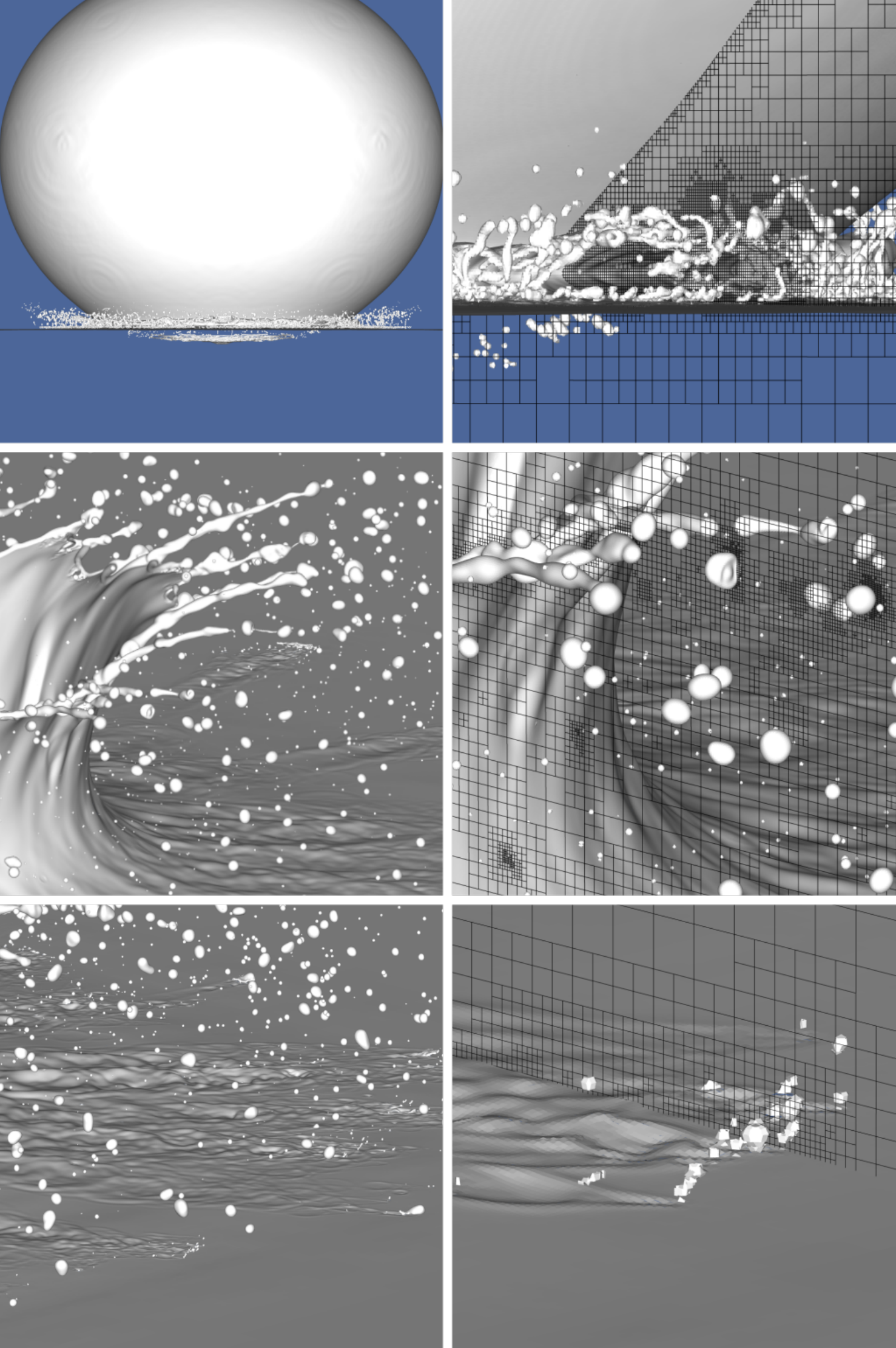} % Images in 100% size
  \put(-3,98.5){\small(\textit{a})}
  \put(-3,65){\small(\textit{b})}
  \put(-3,31){\small(\textit{c})}
  \end{overpic}
  \caption{Different mechanisms of droplet production at different stages of impact. The images are shown under different magnifications. (\textit{a}) $t=30$ $\upmu$s, the ``prompt splash" that occurs at the very early time of the impact near the neck region due to irregular rupture and breakup of the ejecta film. (\textit{b}) $t=650$ $\upmu$s, the sustained ``crown splash" due to breakup of the thin liquid ligaments on the top of the crown rim. (\textit{c}) Partially resolved tiny droplets near the pool surface produced by secondary impact/bubble bursting. The left panel shows the locations of splashing and the right panel demonstrates the mesh structures nearby.}
  \label{fig:dropsource}
\end{figure}

Figure \ref{fig:dropsource} qualitatively illustrates some primary mechanisms of droplet production during the impact captured by simulation. The left panel shows the stages and locations where tiny droplets are generated and the right panel demonstrates the mesh structures on the overlaid section. As shown in figure \ref{fig:dropsource}(\textit{a}), a great number of microdroplets are generated near the neck region immediately after the first contact due to the very early breakups of ejecta film, so-called the ``prompt splash" \citep{deegan2007complexities, marcotte2019ejecta}. We have estimated that around $85\%$ of droplets produced at this time step have the equivalent mean diameter of more than double the size of the smallest cell and $70\%$ of them are in the size range of $10\sim30$ $\upmu$m, suggesting that the statistical analysis of the initial smallest droplets should be treated carefully. After the intricate early splashing, a ``smooth" liquid sheet rises around the contact line to form the cylindrical crown, emanating thin liquid ligaments on its rims from a fairly regular distance. A sustained droplet ejection, so-called the ``crown splash", is therefore observed. The size of droplets produced at this stage is greatly influenced by the thickness of ligaments and increases with time. Figure \ref{fig:dropsource}(\textit{b}) shows the fragmentation on the tips of the ligaments at $t=650$ $\upmu$s. From the right panel, it can be clearly seen that the droplets produced at the present time instant are sufficiently larger than the minimum scale of the cell. Figure \ref{fig:dropsource}(\textit{c}) shows the production of smaller droplets from the ``secondary impact" caused by the previous generation. When the first-born droplets fall back and impinge to the pool, it may produce another splashing or breakups, which are generally partially resolved. The radii of these smallest droplets are represented approximately by the smallest size of the mesh (right panel). Besides the above sources, very small child-drops can be also ejected from bubble bursting, which are usually not fully resolved as same as figure \ref{fig:dropsource}(\textit{c}). Fewer larger droplets are also observed later from the downward-moving central jet after the closure of the upper crown. 

\subsection {Droplet statistics}\label{subsec:dropstatis}
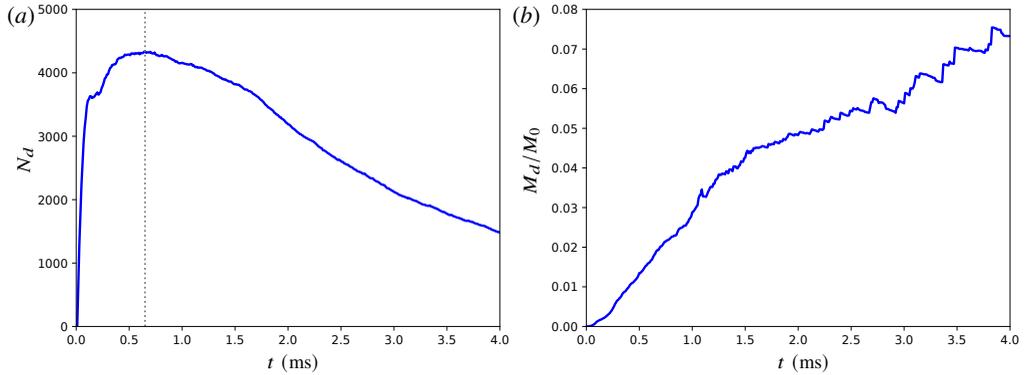
\begin{figure}
  \centering
  \begin{overpic}[width=\linewidth]{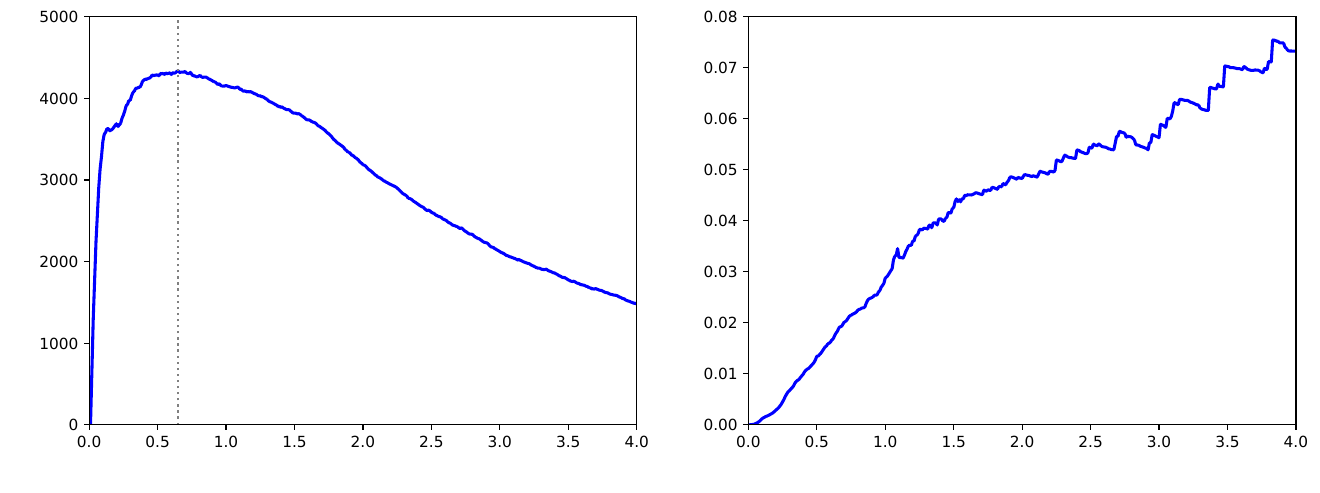} % Images in 100% size
  \put(0,34.7){\small(\textit{a})}
     \put(25.5,1){\scriptsize\textit{$t$ $\rm{(ms)}$}}
     \put(1,20){\scriptsize\rotatebox{90}{\textit{$N_d$}}}
  \put(49.5,34.7){\small(\textit{b})}
     \put(75.5,1){\scriptsize\textit{$t$ $\rm{(ms)}$}}
     \put(50.5,18){\scriptsize\rotatebox{90}{\textit{$M_d/M_0$}}}
  \end{overpic}
  \caption{(\textit{a}) Temporal evolution of the total number of secondary droplets. The vertical dotted line indicates the maximum droplet count at $t\approx650$ $\upmu$s. (\textit{b}) Temporal evolution of the total mass of secondary droplets. The time point when the impacting drop touches the pool is chosen as the reference time. The total mass of secondary droplets $M_d$ is scaled by the mass of the impacting drop $M_0$.}
  \label{fig:dropnum}
\end{figure}

We now discuss the statistics of droplets. Here we only analyze the droplet information captured with higher resolutions ($L_{max}$=13 and 14) for the first 4 ms after impact. The time variation of the total number of droplets is plotted in figure \ref{fig:dropnum}(\textit{a}). It should be noted that a short time disturbance is presented during $t\approx0.21\sim0.27$ ms on the curve, which might be explained as the loss of the smallest droplets at $L_{max}$=14 within the extra refinement layer near the surface of the pool (see section \ref{subsec:congif_spatial}), as it happens at approximately the same time when many droplets reach this height and the extra refinement layer is removed. Neglecting the artefacts at the early time, a unimodal curve of droplet production is found. Shortly after the first contact, a large number of very fine droplets are scattered from the ruptured liquid ejecta (corresponding to figure \ref{fig:dropsource}\textit{a}), which is reflected in figure \ref{fig:dropnum}(\textit{a}) for the sharp increase of the droplet count at the beginning. Subsequently, larger droplets are emitted almost continuously from the tips of the ligaments. The maximum droplet count is reached 650 $\upmu$s after impact with around 4340 droplet population, and the dynamics around this time are qualitatively shown in figure \ref{fig:dropsource}(\textit{b}). As the crown grows, ligaments merge along the rim and become shorter and thicker, producing larger droplets till the closure of the upper part. A change of the decreasing slope can be found at $t\approx$ 1.7 ms in figure \ref{fig:dropnum}(\textit{a}), which indicates the timing when lots of droplets start to exit the field of view and are removed from the computational domain.

Figure \ref{fig:dropnum}(\textit{b}) plots the time evolution of the mass ratio of the total secondary droplets ($M_d$) to the impact drop ($M_0$). An overall increasing trend of the mass transfer from pool to air is found. The number of droplets starts to decrease after the peak ($t\approx650$ $\upmu$s) but the total mass of droplets keeps increasing, which also reveals the fact that the droplets produced at the later ``crown splash" stage are in much larger scales than the early splashing.

\begin{figure}
  \centering
  \begin{overpic}[width=\linewidth]{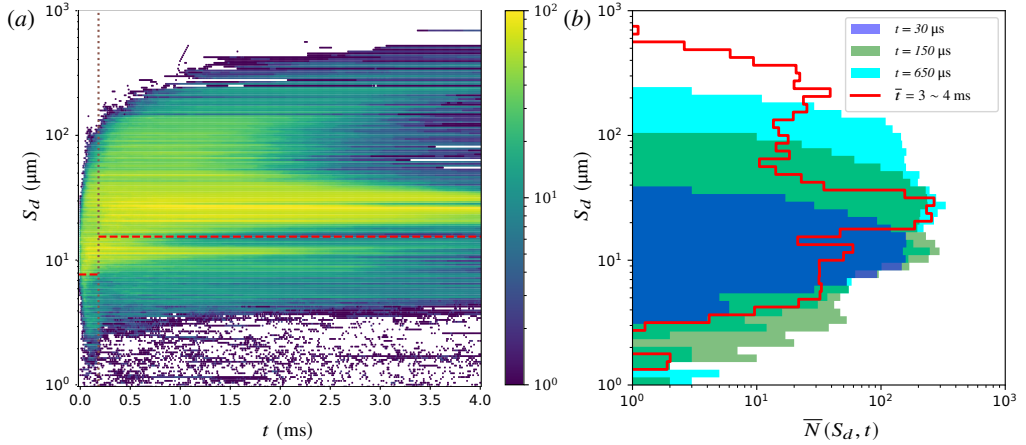} % Images in 100% size
  \put(0,41){\small(\textit{a})}
     \put(25,1){\scriptsize\textit{$t$ $\rm{(ms)}$}}
     \put(1.5,22){\scriptsize\rotatebox{90}{\textit{$S_d$ $\rm{(\upmu m)}$}}}
  \put(54.5,41){\small(\textit{b})}
     \put(87,40.3){\tiny\textit{t = 30 $\rm{\upmu s}$}}
     \put(87,38.2){\tiny\textit{t = 150 $\rm{\upmu s}$}}
     \put(87,36.){\tiny\textit{t = 650 $\rm{\upmu s}$}}
     \put(87,33.8){\tiny\textit{$\overline{t}=3\sim4$ $\rm{ms}$}}
     \put(78,1){\scriptsize\textit{\textit{$\overline{N}(S_d,t)$}}}
     \put(55.5,22){\scriptsize\rotatebox{90}{\textit{$S_d$ $\rm{(\upmu m)}$}}}
  \end{overpic}
  \caption{(\textit{a}) Temporal contour of droplet size distribution during $0\sim4$ ms of impact. The horizontal dashed lines (red) indicate the length scale of doubling the minimum cell size $\sim2\Delta$. (\textit{b}) Droplet size distribution at different time slices in (\textit{a}). Here it shows $t=30$ $\upmu$s and $t=150$ $\upmu$s when the liquid ejecta ruptures and produces very tiny droplets; $t=650$ $\upmu$s when the droplet count reaches the maximum. Large size droplets ($S_d>100$ $\upmu$m) are only generated from the ligament breakups at the ``crown splash" stage. Time-averaged droplet size distribution is calculated using 9 time slices from the time window $t=3\sim4$ ms. A bi-model distribution of droplet size is found.}
  \label{fig:dropsize}
\end{figure}

Figure \ref{fig:dropsize}(\textit{a}) plots the contour map of droplet size distribution per bin size $\Delta r$ at various times. The equivalent mean diameter of each droplet $S_d$ is calculated as a sphere using the integrated volume fraction of the liquid phase. Initially, a lot of relatively small droplets, $S_d$ ranging around $8\sim25$ $\upmu$m, are generated immediately upon impact due to the mechanism of ``prompt splash" (see figure \ref{fig:dropsource}\textit{a}). In the experiments, a ring of fine spray containing mostly $6\sim19$ $\upmu$m droplets was also recorded at the instant of impact, which agrees well with our numerical results, but the existence of smaller droplets was not able to be determined owing to limited camera resolution. In the simulation, a sharp increase of much smaller droplets ($S_d\leqslant8$ $\upmu$m) is captured at the initial splashing stage ($t<200$ $\upmu$s), which can be associated with the partially resolved very small droplets from the rupture and breakup of the early ejecta film and ``secondary impact". It is worth noting that the minimum cell scales are $\Delta\approx3.9$ $\upmu$m at stage $S1$ and $\Delta\approx7.8$ $\upmu$m at stage $S2$ in the present simulation as introduced in \ref{subsec:congif_spatial}, and we assume that the geometries can not be represented for those droplets whose radii are smaller than the minimum cell size (red dashed line in figure \ref{fig:dropsize}\textit{a}), which could influence the breakup physics at smaller scales. However, grid convergence of the numerical atomization remains an open question as has been discussed by \citet{herrmann2013simulating} and \citet{pairetti2020mesh}, although the latter author suggested that 8 cells per drop diameter would probably be enough to preserve most of their physical behaviours. Referring to the recent works of \citet{wang2016high} and \citet{mostert2022high} concerning droplet spray in the action of breaking waves, droplets radii greater than $2\Delta$ would be able to maintain a spherical shape and are approximately grid-converged with proper time-averaging procedures, suggesting that at least $4\sim8$ points per droplet diameter are essential to obtain a numerical convergence. 

After the initial splashing, an ``intact" liquid sheet rises cylindrically from the contact line and disseminates secondary droplets from the thin liquid ligaments as shown in figure \ref{fig:dropsource}(\textit{b}). The size of droplets increases in time with the gradually thickened ligaments at this period. In figure \ref{fig:dropsize}(\textit{a}), three main size concentrations become pronounced 2 ms after impact, which can be briefly classified as the small-sized class $S_d\leqslant16$ $\upmu$m, the medium-sized class $16<S_d<50$ $\upmu$m and the large-sized class $S_d\geqslant100$ $\upmu$m. The first concentration range is associated with the partially resolved smallest droplets and their count decreases noticeably in time, indicating that they are mainly generated at the very beginning of the impact and lots of them fall back to the bulk very quickly after birth. The second concentration range lies in $16\sim50$ $\upmu$m (moderate size) and they are the ones produced the most throughout the entire simulation. The last concentration range is composed of large droplets (larger than 100 $\upmu$m), increasing and narrowing in time, which follows the experimental observation \citep{murphy2015splash} that fewer but larger droplets will be detached from the gradually thickened ligaments with time.

Various Instantaneous time slices of droplet size distribution are provided in figure \ref{fig:dropsize}(\textit{b}). At $t\approx150$ $\upmu$s, the ``bumping" event between the ejecta and the drop/pool just stopped and the primary mechanism of droplet production starts to transition from ``prompt splash" to ``crown splash". Comparing the size distribution at $t=150$ $\upmu$s and $t=650$ $\upmu$s, very similar profiles can be found in the group of moderate droplets ($16\sim50$ $\upmu$m), which confirms that the medium-sized concentration in figure \ref{fig:dropsize}(\textit{a}) is composed primarily of the droplets from the mechanism (\romannumeral2) summarized in \citet{deegan2007complexities}: early rim instability from the very thin ejecta sheet. The discrepancy in small-sized droplets between these two time steps is probably caused by the change of mesh resolution, since the smallest cell size at $t>200$ $\upmu$s has been doubled. Large-sized droplets of $S_d>100$ $\upmu$m appear at $t>150$ $\upmu$s, suggesting that large scale droplets are only produced by the fragmentation of thin ligaments at the ``crown splash" stage.

As the ligaments merge and thicken, the newly detached droplets increase gradually in size, and the most abundant droplet information occurs around $3\sim4$ ms after impact \citep{murphy2015splash}. The time-averaged droplet size distribution over the time window $t=3\sim4$ ms is therefore calculated. As shown in figure \ref{fig:dropsize}(\textit{b}), a bimodal distribution of droplet size can be observed from the histogram, corroborating the separated size concentration ranges in figure \ref{fig:dropsize}(\textit{a}). The bimodal size distribution of liquid sheet fragmentation has been also reported by some other studies \citep{afeti1990distribution, villermaux2009single, villermaux2011drop, murphy2015splash}. Two primary distribution ranges can be briefly distinguished. The first one is composed of small-sized and medium-sized droplets peaking around 30 $\upmu$m. The second primary range appears for the large scale droplets ranging from 100 $\upmu$m to 500 $\upmu$m. 

\begin{figure}
  \centering
  \begin{overpic}[width=\linewidth]{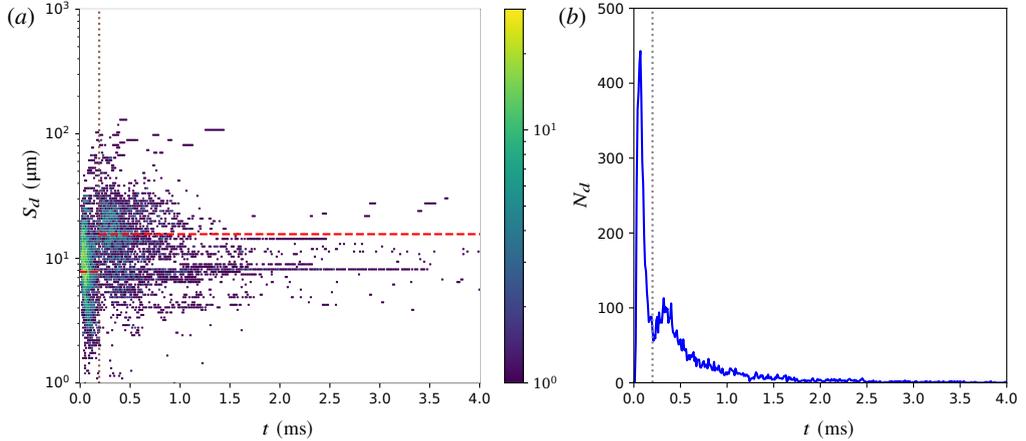} % Images in 100% size
  \put(0,41){\small(\textit{a})}
     \put(25,1){\scriptsize\textit{$t$ $\rm{(ms)}$}}
     \put(1.5,22){\scriptsize\rotatebox{90}{\textit{$S_d$ $\rm{(\upmu m)}$}}}
  \put(54,41){\small(\textit{b})}
     \put(78,1){\scriptsize\textit{$t$ $\rm{(ms)}$}}
     \put(55.5,23){\scriptsize\rotatebox{90}{\textit{$N_d$}}}
  \end{overpic}
  \caption{Statistics of droplets who tend to re-merge with the liquid bulk. (\textit{a}) Temporal contour of the droplet size distribution for the ``re-merging" droplets. (\textit{b}) Evolution of the ``re-merging" droplets count with time.}
  \label{fig:dropreturn}
\end{figure}

Lastly, we would like to focus on those droplets that tend to fall back to the liquid bulk, which is vital for the estimation of mass transfer through the air-sea interface. In practice, those droplets that move along the impact direction (downwards) and whose centroid are located less than 100 $\upmu$m to the free surface of the liquid pool are selected and assumed as the ones that will re-join the pool. Figure \ref{fig:dropreturn}(\textit{a}) shows the temporal contour of the size distribution for the ``re-merging" droplets and (\textit{b}) shows its count in time. It can be found that most of the initially produced small-sized droplets from ``prompt splash" vibrate near the free surface and tend to re-join the bulk shortly after generation. At the beginning of the ``crown splash", very thin ligaments stretch out from the rims of the downward bending ejecta sheet and send out tiny droplets towards the target liquid as demonstrated in figure \ref{fig:valid_interface}(\textit{b}), which thereby contribute to a second primary peak of the medium-sized droplets in figure \ref{fig:dropreturn}(\textit{a,b}). As the impact advances, the direction of the liquid ligaments transients quickly from horizontal-outward to vertical-upward and the droplets are generally pinched off upwards in larger angles accordingly. The amount of the ``re-merging" droplet becomes insignificant 1 ms after impact. Finally, it is worth noting that the presence of wind may significantly advect the smallest droplets and delay or prevent them from rejoining the bulk.

%------------------7 Conclusion---------------------------------------%

\section {Conclusion}\label{sec:conclusion}
In this work, the high-energy splash of drop impact onto a deep volume of the same liquid pool has been investigated with high-resolution direct numerical simulation (DNS) in 3D. The calculations have been conducted in the exact same configurations studied previously by \citet{murphy2015splash}, in order to perform detailed comparisons with the experiments and prepare an in-depth analysis of the splashing dynamics. The BASILISK open-source solver, which combines a VOF description of the gas/liquid interfaces and an adaptive octree grid refinement, has been employed to simulate the process of drop impact. Qualitative and quantitative comparisons between numerical and experimental results have been conducted in terms of the morphological behaviours of splashing, kinematics of crown and cavity as well as the distributions of secondary droplets through a particular field of view, which efficiently validate the present numerical strategies and therefore enable the discussion of the internal mechanisms for high-speed drop impact afterwards.

Following the experimental observations of \citet{murphy2015splash}, we performed a detailed investigation on flow physics and splashing behaviours of high-speed drop impact, serving as an important supplementary study for this issue. Firstly, the very early instantaneous motions in the vicinity of the neck region were discussed under sufficient time resolution. We have confirmed the existence of two different mechanisms of air entrapment in the neck of connection, namely the entrapment of axisymmetric bubble rings driven by high localized pressure at $R_n/R_d<35\%$ \citep{oguz1989surface} and the entrapment of isolated bubbles/bubble rings due to the unstable oscillations of the ejecta base \citep{weiss1999single}. Moreover, the calculation successfully captured the initialization of the irregular azimuthal undulations along the outer edge of the neck, which thereby breaks the axisymmetry of the motions in microscopic scales. Thereafter, detailed information on the internal flows such as velocity, pressure as well as energy budget was extracted from the calculation, to further explain the corresponding physical phenomena observed in experiments. We showed that azimuthal destabilization occurs on the edges of the flattening drop, breaking up on its tips and participating in the production of child-droplets. These ``liquid threads", emitted from the initial impact drop, grow together with the uprising crown and meet upon closure, and they are eventually transported backwards to the bulk along with the penetration of the central spiral jet. Lastly, we presented the statistics of airborne droplets produced by the high-energy drop-pool collision. The results showed that a great number of microdroplets are produced immediately after contact by the irregular ``prompt splash" within the time scale of $t<200$ $\upmu$s, composing the most populated small and moderate sizes. The earliest tiny droplets are sprayed right above the surface of the pool and most of them move towards the target liquid, suggesting the great possibility that a big part of them may return to the liquid bulk shortly after birth. Large droplets ($S_d>100$ $\upmu$m) are only observed from the fragmentation of the rim ligaments during the ``crown splash" stage, which therefore forms a gradually increasing but narrowing distribution ribbon, reflecting the merging and thickening activities of the ligaments on the rim of the crown. Finally, the bimodal size distribution of secondary droplets has been found.   

We are aware that the splashing dynamics induced by the most energetic regime of drop impact are way more complicated than what has been mentioned here. Further analysis of the multi-scale flow physics needs to be conducted under sufficient spatial and temporal resolutions in the future. Details of the bubble ring entrapment are still not well understood, as well as its motions and breakups. The physical mechanism that is responsible for the early azimuthal instability remains unknown. Recent experimental observations \citep{thoraval2013drop, li2018early} have suggested that these early dynamics may be greatly affected by the intricate vortical motions and three-dimensional instabilities, which certainly added more complexities to the analysis. For drop impact on the same liquid pool, large parameter space of $Re$ and $We$ need to be studied to determine the critical conditions of the large bubble entrapment (bubble canopy regime), as well as to explore its influence on the closure event (closure time, closure height, large bubble volume, floating bubble and its burst). The formation and motion of the central spiral jet also need to be analyzed in detail and compared with various types of liquid column jets. Statistics of droplets and bubbles need to be collected under various impacting conditions to form a more comprehensive database, which could therefore reflect directly the mass/momentum transfer through gas-liquid interfaces as well as enlighten various applications where this process is involved (oil dispersant, spray cooling, metallurgy, etc). 
%------------------Acknowledgments------------------------------------%

\backsection[Acknowledgements]{We appreciate beneficial discussions and help from the BASILISK community. Simulations were performed using computational resources on Advanced Research Computing (ARC) at Virginia Tech.}

\backsection[Funding]{This work is supported by the scholarship from China Scholarship Council (CSC) under the Grant CSC NO. 201908320462.}

\backsection[Declaration of interests]{The authors report no conflict of interest.}

\backsection[Author ORCID]{Hui Wang, https://orcid.org/0000-0001-9733-0150; Shuo Liu, https://orcid.org/0000-0002-8530-8359.}
%------------------Appendix------------------------------------------%

\appendix

\section{Effects of spatial resolution}\label{appA}
\begin{figure}
  \centering
  \begin{overpic}[scale=1.5]{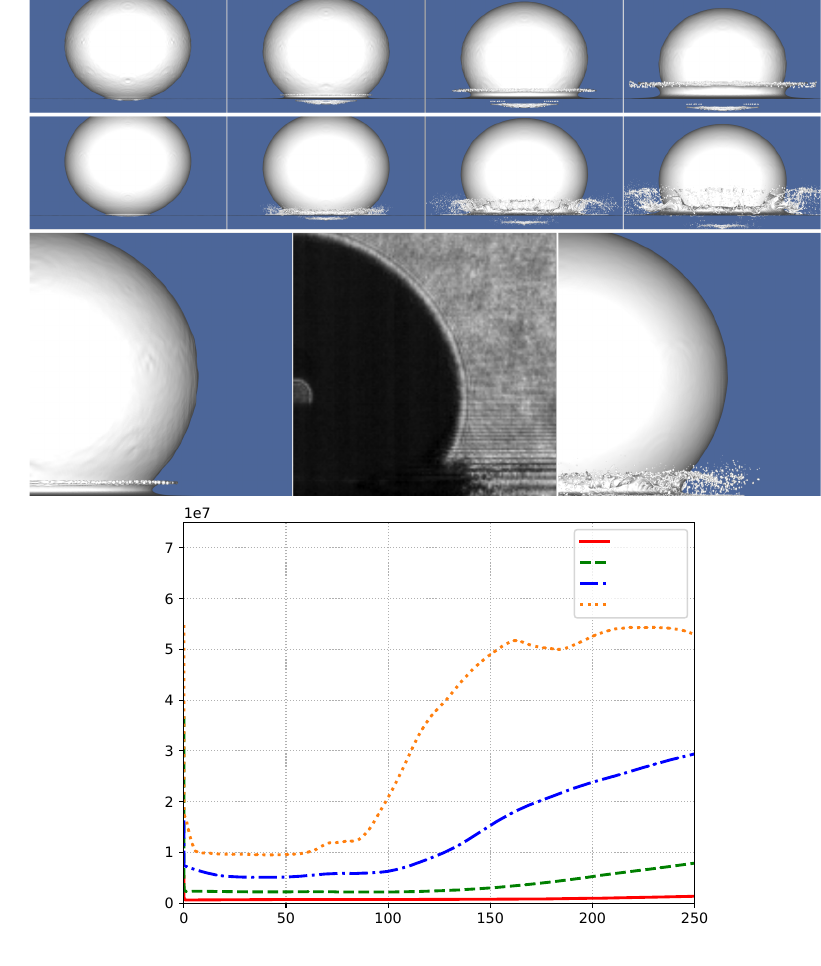} % Images in 100% size
  \put(0,98.4){\small(\textit{a})}
  \put(0,86.2){\small(\textit{b})}
  \put(0,74.3){\small(\textit{c})}
  \put(13,44){\small(\textit{d})}
     \put(64,43.7){\tiny\textit{$L_{max}$ = $11$}}
     \put(64,41.5){\tiny\textit{$L_{max}$ = $12$}}
     \put(64,39.4){\tiny\textit{$L_{max}$ = $13$}}
     \put(64,37.2){\tiny\textit{$L_{max}$ = $14$}}
     \put(43,2){\scriptsize\textit{$t$ $(\upmu\rm {s})$}}
     \put(14,25){\scriptsize\rotatebox{90}{\textit{$N_c$}}}
  \end{overpic}
  \caption{Effects of the maximum mesh refinement level on splashing behaviours. (\textit{a}) Evolution of early-time splashing obtained with $L_{max}=12$, showing 10, 40, 70 and 100 $\upmu$s after impact. (\textit{b}) Evolution of early-time splashing obtained with $L_{max}=14$, showing 10, 40, 70 and 100 $\upmu$s after impact. (\textit{c}) Comparison of air-water interfaces at $t=50$ $\upmu$s captured by calculation with $L_{max}=12$ (left), experiment (middle) and calculation with $L_{max}=14$ (right). (\textit{d}) Time evolution of  cell numbers under different maximum mesh refinement levels.}
  \label{fig:appenA_Lmax}
\end{figure}
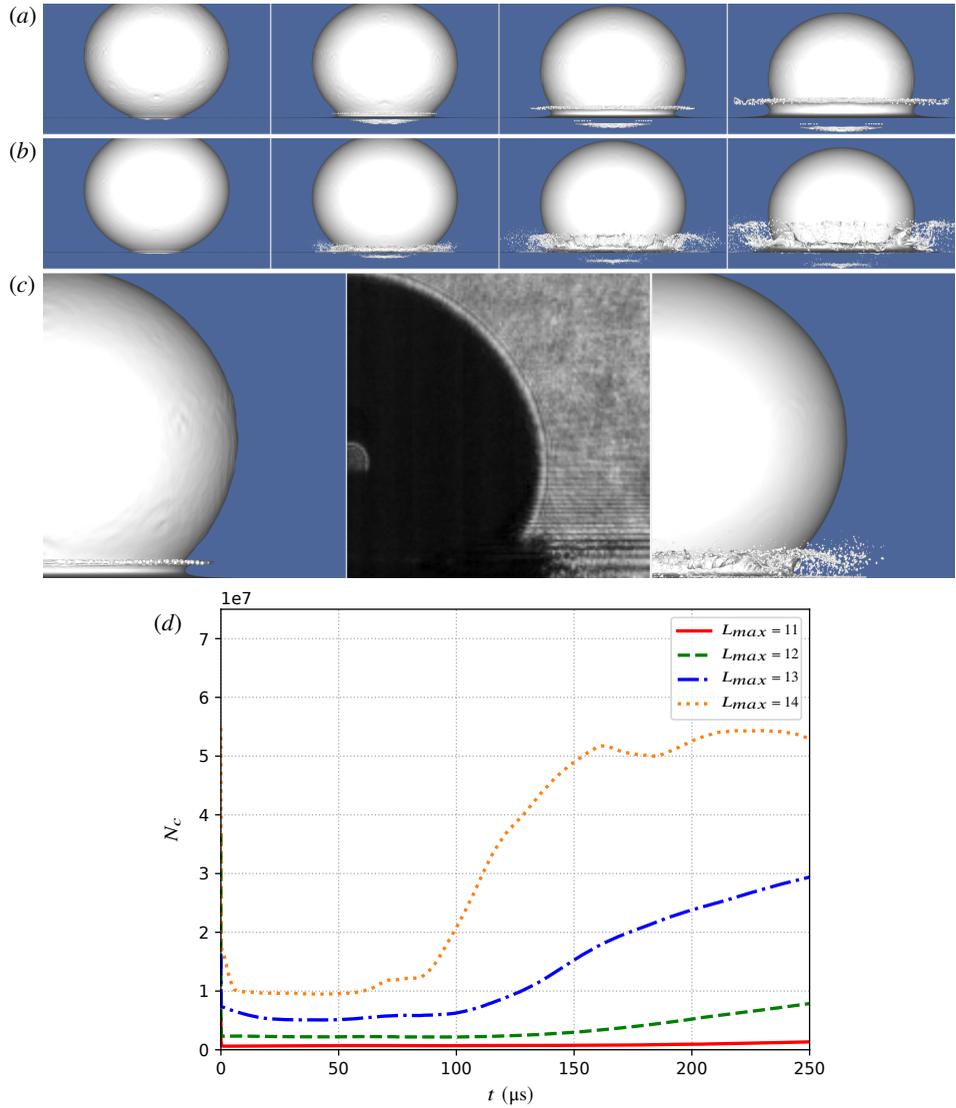

In this appendix, we discuss the effects of the minimum spatial resolution. Using the Adaptive Mesh Refinement (AMR) algorithm, preliminary tests have been carried out by varying the maximum refinement level $L_{max}$ from 11 to 14, corresponding to the minimum cell size of $3.9\sim31.25$ $\upmu$m in reality. Two types of distinctive splashing phenomena are observed under different spatial resolutions. For simulations at lower levels ($L_{max}$=12 in figure \ref{fig:appenA_Lmax}\textit{a}), a thin layer of ejecta sheet rises smoothly from the contact line and later becomes the leading edge of the crown, and secondary droplets detach nearly axisymmetrically from the rim of the liquid sheet. Increasing the maximum mesh refinement to higher levels ($L_{max}$=14 in figure \ref{fig:appenA_Lmax}\textit{b}), a disturbed incoherent liquid sheet accompanied by more ``randomly" distributed secondary droplets are captured (details in section \ref{subsec:early} and \ref{sec:droplet}). Referring the splashing classifications in \citet{thoroddsen2011droplet} and \citet{thoraval2012karman} based on the dimensionless Ohnesorge number ($Oh=\mu/\rho\sigma d$) and Splash number ($K=We\sqrt{Re}$), irregular splashing always occurs for the most energetic cases. As the parameters in our present case are much higher than their range of study, thinner liquid sheet and more complex interfacial deformations/breakups can be even expected. 

Figure \ref{fig:appenA_Lmax}(\textit{c}) compares the shape of the air-water interface captured with $L_{max}$=12 (left), high-speed camera (middle) and $L_{max}$=14 (right) 50 $\upmu$s after impact. The emerge-ruptured liquid sheet together with a great number of irregularly distributed tiny droplets calculated at $L_{max}$=14 are consistent well with the experimental observations, while a rather smooth ejecta is presented with $L_{max}$=12 when the spatial resolution is insufficient. As a formal study of mesh influence could not be performed, given the computational resources required, the successful reproduction of the primary features of the early splashing confirms that a maximum refinement level at $L_{max}=14$ is essential for capturing the ``correct" physical dynamics (see also section \ref{subsec:valid_morph}).

It should be noted that each time the maximum refinement level is increased, eight ``children" cells will be divided from the tree-based structure in 3D. Although the adaptive wavelet algorithm concentrates the smallest scales mostly near the interfaces of the impacting area, the computational requirements remain harsh at high levels. Figure \ref{fig:appenA_Lmax}(\textit{d}) shows the time evolution of the cell number for the first 250 $\upmu$s of the simulations under different spatial resolutions. It can be observed that the total number of cells is more than double with each additional increase of the $L_{max}$, and this number will become considerably huge later when the large crater and the substantial droplets and bubbles are developed.
  
%------------------Reference------------------------------------------%

\bibliographystyle{jfm}
% Note the spaces between the initials
\bibliography{jfm}

\end{document}